\documentclass[smallextended]{svjour3}

\usepackage[utf8]{inputenc}
\usepackage{mathpazo}
\usepackage{amsmath}
\usepackage[numbers]{natbib}
\usepackage[margin=1in]{geometry}
\usepackage{amssymb}
\usepackage{standalone}
\usepackage{graphicx}
\usepackage{epstopdf}
\usepackage[normalem]{ulem}
\usepackage[rgb,dvipsnames]{xcolor}
\usepackage{tikz}
\usepackage{bm}
\usepackage[caption=false]{subfig}
\usepackage{booktabs}
\AtBeginDocument{
    \heavyrulewidth=.08em
    \lightrulewidth=.05em
    \cmidrulewidth=.03em
    \belowrulesep=.65ex
    \belowbottomsep=0pt
    \aboverulesep=.4ex
    \abovetopsep=0pt
    \cmidrulesep=\doublerulesep
    \cmidrulekern=.5em
    \defaultaddspace=.5em
}

\definecolor{MyBlue}{rgb}  {0.1,0.1,0.9}
\definecolor{MyRed}{rgb}   {0.9,0.1,0.1}
\definecolor{MyGreen}{rgb} {0.05,0.4,0.05}
\definecolor{burntorange}{rgb}{0.8, 0.33, 0.0}
\definecolor{NeilMagenta}{rgb}{0.8, 0.1, 0.8}

\newcommand \bk{\color{black}}
\newcommand \gr{\color{MyGreen}}

\begin{document}
\title{Turing Patterning in Stratified Domains}
\titlerunning{Turing Patterning in Stratified Domains}
\author{Andrew L. Krause \and V\'{a}clav Klika \and Jacob Halatek \and Paul K. Grant \and Thomas E. Woolley \and Neil Dalchau \and Eamonn A. Gaffney}
\authorrunning{A. L. Krause \and V. Klika \and J. Halatek \and P. K. Grant \and T. E. Woolley \and N. Dalchau \and E. A. Gaffney}

\institute{A. L. Krause \and E. A. Gaffney 
    \at Wolfson Centre for Mathematical Biology, Mathematical Institute, University of Oxford, Andrew Wiles Building, Radcliffe Observatory Quarter, Woodstock Road, Oxford, OX2 6GG, United Kingdom 
\and
V. Klika 
    \at Department of Mathematics, FNSPE, Czech Technical University in Prague, Trojanova 13, 120 00 Praha, Czech Republic
\and
J. Halatek \and P. K. Grant \and N. Dalchau 
    \at Microsoft Research, 21 Station Rd, Cambridge, CB1 2FB, UK
\and
T. E. Woolley 
    \at Cardiff School of Mathematics, Cardiff University, Senghennydd Road, Cardiff, CF24 4AG, United Kingdom}

\date{Received: date / Accepted: date}

\maketitle

\begin{abstract}
   Reaction-diffusion processes across layered media arise in several scientific domains such as pattern-forming \emph{E. coli} on agar substrates, epidermal-mesenchymal coupling in development, and symmetry-breaking in cell polarisation. We develop a modelling framework for bi-layer reaction-diffusion systems and relate it to a range of existing models. We derive conditions for diffusion-driven instability of a spatially homogeneous equilibrium analogous to the classical conditions for a Turing instability in the simplest nontrivial setting where one domain has a standard reaction-diffusion system, and the other permits only diffusion. Due to the transverse coupling between these two regions, standard techniques for computing eigenfunctions of the Laplacian cannot be applied, and so we propose an alternative method to compute the dispersion relation directly. We compare instability conditions with full numerical simulations to demonstrate impacts of the geometry and coupling parameters on patterning, and explore various experimentally-relevant asymptotic regimes. In the regime where the first domain is suitably thin, we recover a simple modulation of the standard Turing conditions, and find that often the broad impact of the diffusion-only domain is to reduce the ability of the system to form patterns. We also demonstrate complex impacts of this coupling on pattern formation. For instance, we exhibit non-monotonicity of pattern-forming instabilities with respect to geometric and coupling parameters, and highlight an instability from a nontrivial interaction between kinetics in one domain and diffusion in the other. These results are valuable for informing design choices in applications such as synthetic engineering of Turing patterns, but also for understanding the role of stratified media in modulating pattern-forming processes in developmental biology and beyond.
\end{abstract}
 

\section{Introduction}
Since Turing's initial insights into reaction-diffusion driven morphogenesis \cite{turing1952chemical}, a substantial research effort has elucidated various mathematical and biophysical aspects of such symmetry-breaking instabilities leading from homogeneity to patterned states \cite{de1991turing, cross1993pattern, maini2012turing, kondo2010reaction, green2015positional,woolley2014visions}. An important and well-studied aspect of these instabilities is the underlying geometry, which can influence both the stability of a homogeneous state, as well as the subsequent mode selection of emergent patterns \cite{Murray2003}. However, one less well-studied aspect of geometry is the coupling between layered spatial domains, which can arise in a variety of settings and is the primary object of interest in this paper.

Reaction diffusion processes arise in a diversity of layered settings, from bulk-surface membrane-cytosol interactions \cite{halatek2018review, ratz2014symmetry, kretschmer2016pattern, spill2016effects, cusseddu2018coupled} to epithelial-mesenchymal couplings in developing skin \cite{cruywagen1992tissue, shaw1990analysis}. Synthetic experiments involving pattern formation in monolayers also exhibit clearly stratified regions of cells and culture medium \cite{sekine2018synthetic}. Additionally, many experiments involving bacterial pattern-formation are performed using colonies on the surface of a substrate, such as agar \cite{budrene1991complex, budrene1995dynamics}. Such systems either use natural chemotaxis mechanisms to initiate spatial pattern formation of the bacterial density itself \cite{tyson1999minimal}, or instead use synthetic bacteria re-engineered to express additional quorum-sensing pathways that spatially coordinate patterns in gene expression \cite{Basu2005bandpass, Tabor2009edge, grant2016orthogonal}. Other examples are synthetically reconstituted protein interaction systems with bulk-membrane coupling such as the Min system \cite{loose2008exp, kretschmer2016pattern, frey2018review}, where molecular interactions \cite{denk2018syth,glock2019synth}, or {\it in vitro} system geometries \cite{wu2016geometry, brauns2020bulk, halatek2018box}, are modified to stimulate changes in the observed protein patterns. 
Examples of particular contemporary interest include the use of bacterial colonies as exemplars of synthetic multicellular communication and self-organisation \cite{balagadde2008synthetic, dalchau2012towards, Payne2013rings, grant2016orthogonal, karig2018stochastic}, for example using modified \emph{E. coli} with engineered quorum-sensing signalling on the surface of an agar plate \cite{grant2016orthogonal, Payne2013rings, Cao2016scale,Cao2017pressure}. Some of these systems take advantage of the geometry of colony growth and nutrient diffusion to influence pattern formation \cite{Payne2013rings,Cao2016scale,Cao2017pressure} while in other systems the bacteria are confined \cite{grant2016orthogonal, boehm2018}, but the signalling molecules can diffuse into the inert agar layer below the chemically active colonies. The impact of this leaching on the prospects of a Turing instability in experimentally-relevant geometries has not been fully characterised and is a key motivation for our study.

Our first objective is to develop a two-domain model of reaction-diffusion processes coupled in a stratified bi-layer and to determine conditions for the Turing instability, on the assumption that the upper region is sufficiently substantive in the transverse direction to merit continuum modelling. Such a model is also applicable to a variety of other settings beyond multi-layered bacterial pattern formation, such as developing skin. Our second objective is to focus on the Turing instability for multi-layered bacterial systems, where signalling molecules only diffuse in the lower (agar) layer and especially where the upper layer is asymptotically thin relative to the scale of the pattern and the depth of the lower layer. The main biological motivation is to determine to what extent the diffusive bulk helps, or hinders, the ability of an engineered system to exhibit Turing-type patterning.

In terms of model development, domain-coupled reaction-diffusion systems broadly fall into three major types: instantaneously coupled, bulk-surface models and bulk-bulk models. The first type are models where the components occupy the same physical space (or the reactions occur in thin regions where a homogenisation approximation is sensible) \cite{yang2002spatial, epstein2007coupled, yang2003oscillatory, fujita2013pattern}. Such models are essentially just larger reaction-diffusion systems with linear coupling between subsystems, and amenable to block-matrix analysis in the study of Turing instabilities  \cite{catlla2012instabilities}, but do not capture the spatial separation of the domains. When applied to layered media, these models effectively assume vertical transport between distinct layers (such as that of Figure~\ref{Schematic}) is instantaneous. However, considering physical scales representative of synthetic pattern formation experiments using {\it E. coli} \cite{grant2016orthogonal,boehm2018}, and summarised below in Table~\ref{tab0}, one has an agar block with a depth of a few millimetres, say three, and a diffusion rate on the scale of $4\times 10^{-10}$ m$^2$ s$^{-1}$. Thus the timescale for vertical transport is in the region of 375 minutes, which is short compared to the timescales on which experimental measurements of the equilibrated system are recorded (1500-3000 minutes, \cite{grant2016orthogonal,boehm2018}) but far from instantaneous. 
Hence, such models are inappropriate for the motivating examples here.

A second class of model considers bulk-surface coupling, where one component is confined to the boundary of the main bulk domain, and reactants flow between the two regions, such as in the case of proteins diffusing in the cytoplasm and binding on the cell membrane \cite{ratz2014symmetry, madzvamuse2015stability, spill2016effects, cusseddu2018coupled, paquin2018pattern, halatek2018review, frey2018review}. There is substantial recent interest in such models, from very theoretical results on existence and fast-reaction limiting behaviour \cite{ratz2015turing, anguige2017global, hausberg2018well}, to spike dynamics \cite{gomez2018linear} and a myriad of applications to understanding cell polarity \cite{thalmeier2016geometry, kretschmer2016pattern, halatek2018review, gesele2020ellipse}. One particularly well studied example is the pole-to-pole Min protein oscillation in {\it E. coli}, which has the biological function of guiding the cell division machinery to midcell \cite{kretschmer2016pattern}. Such intracellular protein patterning systems have been studied experimentally and theoretically in a wide range of system geometries, such as spherical \cite{klunder2013sphere, levine2005membrane}, elliptical \cite{halatek2012ellipse, wu2016geometry, gesele2020ellipse}, and planar membrane geometries \cite{halatek2018box}. A striking feature of these examples is that the geometry itself has a major impact on pattern formation and pattern selection, which has been confirmed experimentally \cite{wu2016geometry, brauns2020bulk}.  
More generally, the Turing instability has also been studied in the context of membrane-cytosol models \cite{ratz2014symmetry, madzvamuse2015stability}. 
Overall, linear stability analysis (as used by Turing) is highly applicable to such membrane-cytosol systems because the nonlinear interactions are typically restricted to the lower-dimensional membrane surface. The dynamics in the extended bulk are typically linear such that a general solution (or a good approximation) can be obtained analytically and used to satisfy the linearised reactive boundary condition. This is justified because the membrane can be considered as a surface with no transverse extent, and so transverse gradients only play a role in the cytosolic layer close to the membrane surface.    
However, in multi-layered cellular systems the transverse lengthscales are at least that of many cells and hence transverse gradients cannot be neglected {\it a priori}, and thus should be accommodated in the modelling. 
Models accounting for this represent the final class, with two separated spatial domains with an interface and suitable coupling boundary conditions. From the perspective of pattern formation, this kind of model has only been subject to recent numerical exploration \cite{vilaca2019numerical}, though it is used in the derivation of the second class -- bulk-surface models -- given appropriate distinguished limits and scaling assumptions (for example, \cite{chapman2016reactive, fussell_hybdrid_2018}). 

Hence, we will develop models of the latter type, with an exploration of the conditions for the Turing instability, and their detailed study in the context of a stratified model with an inert underlying agar layer. Turing instabilities of reaction-diffusion systems have been studied on a variety of complex spatial domains such as compact manifolds \cite{varea1999turing, chaplain2001spatio}, networks \cite{asllani2014theory, ide2016turing,nakao2010turing}, and many of the aforementioned complex system geometries \cite{halatek2012ellipse, klunder2013sphere, halatek2018box}. The primary difficulty in such cases, compared to the textbook example of a continuous line, is determining the corresponding set of eigenfunctions and eigenvalues of the spatial transport operators, which for some system geometries do not need to coincide between domains (e.g. in the surface-bulk elliptical case \cite{halatek2012ellipse}). In such cases approximate solutions for the system's eigenfunctions need to be derived that are orthogonal in the patterning layer. 
Examples that deviate even further from the classical case are growing domains \cite{crampin1999reaction, plaza2004effect, KrauseAnisotropy2018, SanchezGarduno2018} and spatially heterogeneous reaction-diffusion processes \cite{benson1998unravelling, page2003pattern, page2005complex, haim2015non, kolokolnikov2018pattern}, for which the canonical approach does not work. In such cases, novel approaches to pattern-forming instabilities have recently been developed for growth \cite{madzvamuse2010stability, van_gorder_growth_2019} and heterogeneity \cite{krause_WKB} under certain simplifications, but such analyses are quite different to the classical case. In a similar direction, as part of our objective in exploring the Turing instability for layered reaction-diffusion systems, we will aim to demonstrate a much richer diversity of structure in the resulting dispersion relations (and hence instability conditions), compared to classical counterparts.

As an outline, in Section~\ref{modelling} we present a two-domain layered model, where each domain consists of  closed two-dimensional rectangular regions,  coupled through a single shared boundary (See Figure~\ref{Schematic}) and briefly discuss how it can be reduced to a variety of other models. We focus on a special case of a two-domain model where we assume linear coupling and no reactions in the second (bulk) region, but note that our analysis can be applied with relatively simple modifications to more general cases. In Section~\ref{GenLinAnalysis} we develop an approach to linear stability analysis of homogeneous states. In Section~\ref{Asymptotics} we derive a variety of asymptotic results regarding our dispersion relation, especially considering limits that are of particular relevance for synthetic pattern formation in {\it E. coli} colonies. We further explore these results and other parameter regimes numerically in Section~\ref{Numerics}. Finally we discuss our results in Section~\ref{Discussion}.

\section{Two-Region Model}\label{modelling}

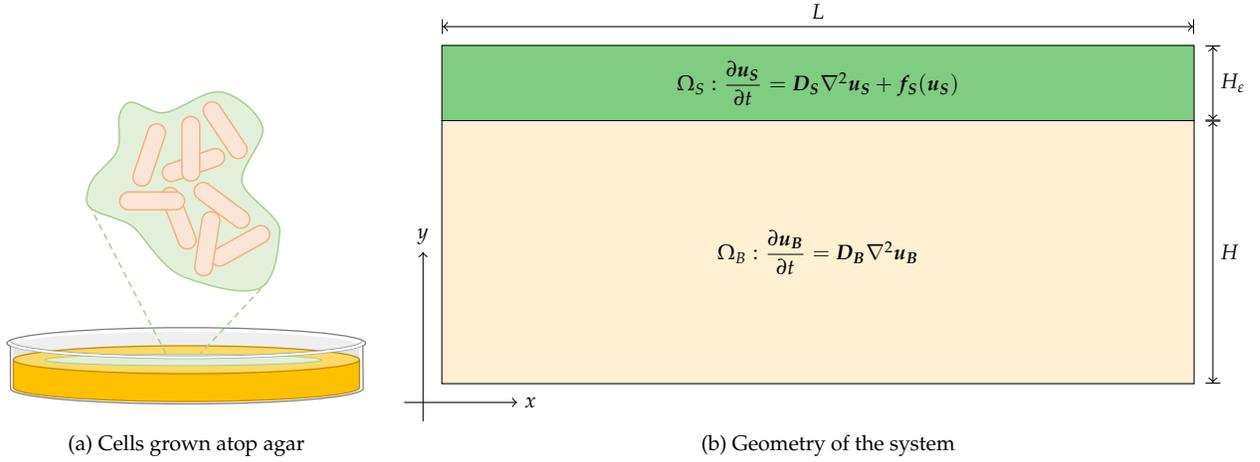
\begin{figure}[ht] \centering
\subfloat[Cells grown atop agar]{\includegraphics[width=0.35\linewidth]{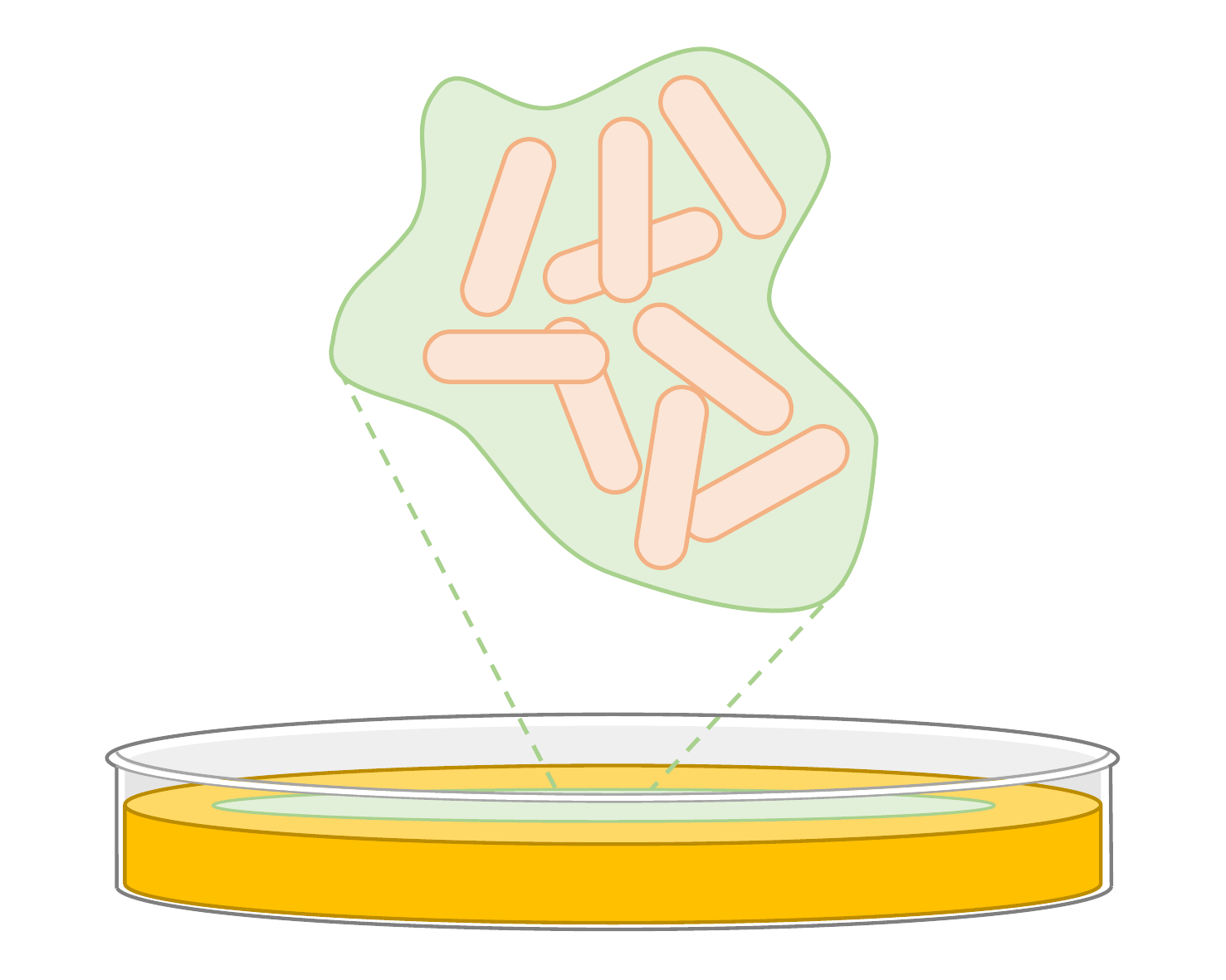}}
\subfloat[Geometry of the system]{\begin{tikzpicture}[scale=5]
\def\H{0.7}
\def\L{2.0}
\def\epsilon{0.2}
\def\dx{0.05}
\def\dy{0.05}

\filldraw[fill=OliveGreen!50!white] (0,\H) rectangle (\L,\H+\epsilon) node[midway] {$\Omega_S: \dfrac{\partial \bm{u_S}}{\partial t} = \bm{D_S}\nabla^2 \bm{u_S}+\bm{f_S}(\bm{u_S})$};

\filldraw[fill=Dandelion!20!white,draw=black] (0,0) rectangle (\L,\H) node[midway] {$\Omega_B: \dfrac{\partial \bm{u_B}}{\partial t} = \bm{D_B}\nabla^2 \bm{u_B}$};

\draw[|<->|](0,\H+\epsilon+\dy) -- (\L,\H+\epsilon+\dy) node[midway, above] {$L$} ;
\draw[|<->|](\L+\dx,0) -- (\L+\dx,\H) node[midway, right] {$H$} ;
\draw[|<->|](\L+\dx,\H) -- (\L+\dx,\H+\epsilon) node[midway, right] {$H_\varepsilon$} ;

\draw[->] (-\dx-\dy,-\dy)--(0.1*\L,-\dy) node[right] {$x$};
\draw[->] (-\dx,-\dx-\dy)--(-\dx,0.5*\H) node[above] {$y$};
\end{tikzpicture}}
\caption{Example experimental system under consideration. (a) Here we consider cells growing in culture on top of a solid reservoir of nutrients, such as agar. (b) The surface (cellular) region denoted $\Omega_S$ has height $H_\varepsilon$ and contains both reaction and diffusion terms, whereas the bulk (nutrient) region $\Omega_B$ is of height $H$ and is assumed to have no reactions, but permits diffusion. Both have lateral extent $L$, with no-flux conditions on all boundaries except for the interface between the two regions, where a coupling condition is applied.\label{Schematic}}
\end{figure}

We consider a layered two-domain model where each domain is governed by a different reaction-diffusion system. We consider several interacting species in these two domains, which we write as $\Omega = \Omega_S \bigcup \Omega_B$ where we refer to $\Omega_B = [0,L]\times [0,H]$ as the bulk region, and $\Omega_S = [0,L]\times [H,H+H_\varepsilon]$ as the surface region (see Figure~\ref{Schematic}). We write $\bm{\hat{u}_B} \in \mathbb{R}^n$ for the concentrations of reactants in the bulk, and $\bm{\hat{u}_S}\in \mathbb{R}^n$ for the concentrations of reactants in the surface region. For simplicity, we consider a simple one-dimensional lateral geometry (orthogonal to the direction of the coupling condition), but note that the geometric details in the lateral direction(s) can be easily extended to much more complicated geometries, as long as eigenfunctions of the Laplacian in these directions are separable from the transverse coordinate $y$.  We only consider reactions on the surface layer and assume the bulk only permits diffusion.

We have the following equations for the species concentrations in the bulk and surface regions:
\begin{equation}\label{eqSdim}
\frac{\partial \bm{\hat{u}_S}}{\partial \hat{t}} = \bm{\hat{D}_S}\nabla^2 \bm{\hat{u}_S}+\bm{\hat{f}_S}(\bm{\hat{u}_S}), \quad \hat{x} \in [0,L], \quad \hat{y} \in [H,H+H_\varepsilon],
\end{equation}
\begin{equation}\label{eqBdim}
\frac{\partial \bm{\hat{u}_B}}{\partial \hat{t}} = \bm{\hat{D}_B}\nabla^2 \bm{\hat{u}_B} , \quad \hat{x} \in [0,L], \quad \hat{y} \in [0,H],
\end{equation}
 where $\bm{\hat{D}_S}, ~\bm{\hat{D}_B}$ are positive definite diagonal matrices. We further specify Neumann (no-flux) boundary conditions on the outer boundaries as,
\begin{equation}\label{BC1dim}
\frac{\partial \bm{\hat{u}_S}}{\partial \hat{x}} = \frac{\partial \bm{\hat{u}_B}}{\partial \hat{x}} = \bm{0}, \text{ for } \hat{x}=0, L,
\end{equation}
\begin{equation}\label{BC2dim}
\frac{\partial \bm{\hat{u}_S}}{\partial \hat{y}} = \bm{0},  \text{ for } \hat{y}=H+H_\varepsilon, \quad \frac{\partial \bm{\hat{u}_B}}{\partial \hat{y}} = \bm{0}, \text{ for } \hat{y}=0,
\end{equation}
and lastly coupling conditions on the interior boundary which conserve fluxes and take the form,
\begin{equation}\label{couplingdim}
\bm{\hat{D}_S}\frac{\partial \bm{\hat{u}_S}}{\partial \hat{y}} = \hat{\eta}\bm{\hat{g}}(\bm{\hat{u}_S}, \bm{\hat{u}_B}), \quad \bm{\hat{D}_B}\frac{\partial \bm{\hat{u}_B}}{\partial \hat{y}} = \hat{\eta}\bm{\hat{g}}(\bm{\hat{u}_S}, \bm{\hat{u}_B}),  \text{ for } \hat{y}=H,
\end{equation}
where $\bm{\hat{g}}$ is a given function determining the transport between the surface and the bulk region, and $\hat{\eta}$ is a rate of transport across the boundary. Essentially, all of the forthcoming analysis can be carried out with a general $\bm{\hat{g}}$, as linearisation will also linearise this function. For brevity and concreteness, we will henceforth assume a linear transport law, so that we have \begin{eqnarray}\label{geqn}
\bm{\hat{g}} = \bm{\hat{u}_S}-\bm{\hat{u}_B}.
\end{eqnarray}

We non-dimensionalise the above model via concentration, time and length scales corresponding to the reaction kinetics and a unit lengthscale $\hat{L}$, respectively. Specifically, we define $\bm{\hat{u}_S} = \bm{U}\bm{{u}_S}$, $\bm{\hat{u}_B} = \bm{U}\bm{{u}_B}$, where $\bm{U}$ is a diagonal matrix of concentration scales. Equally, we set $\hat{t} = \tau {t}$, where $\tau$ is  the timescale of the fastest reaction in the surface and bulk, and $\bm{\hat{x}} = \hat{L} \bm{x}$. The dimensional scalings are then chosen such that 
\begin{eqnarray}\label{sceqn}
\bm{\hat{f_S}}(\bm{\hat{u}_S}) =(1/\tau) \bm{U}\bm{{f}_S}(\bm{{u}_S}), ~~~  \bm{\hat{g}}(\bm{\hat{u}_S},\bm{\hat{u}_B}) =   \bm{U}{\bm{g}}(\bm{{u}_S},\bm{{u}_B}).
\end{eqnarray}
We define new dimensionless groupings $h = H/\hat{L}$, $\varepsilon = H_\varepsilon/\hat{L}$, $\bm{D_S} = \tau\bm{\hat{D}_S}/(\hat{L}^2)$, $\bm{{D}_B} = \tau\bm{\hat{D}_B}/(\hat{L}^2)$, $\tilde{L} = L/\hat{L}$ and $\eta = \tau\hat{\eta}/\hat{L}$. The nondimensional system is written as
\begin{equation}\label{eqS}
\frac{\partial \bm{u_S}}{\partial t} = \bm{D_S}\nabla^2 \bm{u_S}+\bm{f_S}(\bm{u_S}), \quad x \in [0,\tilde{L}], \quad y \in [h,h + \varepsilon],
\end{equation}
\begin{equation}\label{eqB}
\frac{\partial \bm{u_B}}{\partial t} = \bm{D_B}\nabla^2 \bm{u_B}, \quad x \in [0,\tilde{L}], \quad y \in [0,h],
\end{equation}
\begin{equation}\label{BC1}
\frac{\partial \bm{u_S}}{\partial x} = \frac{\partial \bm{u_B}}{\partial x} = \bm{0}, \text{ for } x=0\text{ and } \tilde{L},
\end{equation}
\begin{equation}\label{BC2}
\frac{\partial \bm{u_S}}{\partial y} = \bm{0},  \text{ for } y=h+\varepsilon, \quad \frac{\partial \bm{u_B}}{\partial y} = \bm{0}, \text{ for } y=0,
\end{equation}
\begin{equation}\label{coupling}
\bm{D_S}\frac{\partial \bm{u_S}}{\partial y} = \eta\bm{g}(\bm{u_S}, \bm{u_B}), \quad \bm{D_B}\frac{\partial \bm{u_B}}{\partial y} = \eta\bm{g}(\bm{u_S}, \bm{u_B}),  \text{ for } y=h.
\end{equation}

There are several distinguished limits of the nondimensional system \eqref{eqS}-\eqref{coupling} that reduce the model to different cases already present in the literature. In the limit $\varepsilon \to 0$, one can consider either scaling $\eta \sim O(\varepsilon)$ or scaling $\bm{f_S} \sim O(\varepsilon^{-1})$ in order to reduce the system to a bulk-surface model, which is well-studied in the literature (though primarily in radial geometries) \cite{levine2005membrane, ratz2014symmetry, ratz2015turing, madzvamuse2015stability,cusseddu2018coupled, gomez2018linear, paquin2018pattern}. The second scaling, indicating that the surface timescale is rapid, can be related to assumptions regarding rapid surface reactions used to justify reactive boundary conditions from the microscopic viewpoint \cite{chapman2016reactive}. Finally another limit is the case of infinite permeability, $\eta \to \infty$, wherein the concentrations and fluxes are continuous across the interface. In this case, the system can be seen as a single domain model with a step function heterogeneity, which has been studied extensively as an example of spatially heterogeneous reaction-diffusion systems \cite{benson1998unravelling, page2003pattern, stephetero}. Nonetheless, pattern formation in the system above, as well as several other distinguished limits, has not been analysed yet in the literature. 

In Table~\ref{tab0}, we give the dimensional parameter scales to be considered in our framework, taken from the key motivating example of synthetic patterning in {\it E. coli} bacterial colonies on an agar substrate \cite{grant2016orthogonal,boehm2018}. While such experiments can be conducted with a variety of settings, an overall restriction on the variation of these parameters is motivated by  the range of the physical scales in these studies.  Here, bacteria are plated in squares of about 1 mm (Methods, \cite{boehm2018}) with patterning cells considered in an 8$\times$8 grid in one study (Supplementary Information, \cite{grant2016orthogonal}) and more generally the patterning fields are observed across about 22 such squares  (Fig 5B, \cite{boehm2018} and Fig 3E \cite{grant2016orthogonal}). 
Thus we consider a range of $\hat{L} \sim 8-22$ mm. 
For the diffusion matrices, the infinity (max) norm $\| \cdot \|_\infty$ is presented, i.e. the maximum value of  the matrix's components.   
From Grant et al., (Supplementary Material, Tables S8, S9, \cite{grant2016orthogonal}) diffusion coefficients have been estimated in the range $\| \bm{\hat{D}_S}\|_\infty \sim 10^{-10}$ m$^2$s$^{-1} - 10^{-9}$ m$^2$s$^{-1}$ by model fitting to exemplar results. 
As this is also the scale of diffusion (or slightly more than the scale)  for the signalling molecule EGF in water \cite{diffsize}, the same scale is used for $\| \bm{\hat{D}_B}\|_\infty$.
Similarly, in the parameter fitting by Grant et. al., a reaction timescale on the scale of the faster reactions is such that $1/\tau \sim 8.4\times10^{-5}$ s$^{-1} - 10^{-3}$ s$^{-1}$, with the range arising from the use of different model kinetics in parameter fitting. We further assume 10-50 layers of bacteria, with an {\it E. coli} bacterium size scale of about $10^{-6}$ m$-2\times 10^{-6}$ m \cite{bactsize}, and hence a surface depth on the scale of $H_\varepsilon\sim 10^{-5}$ m$-  10^{-4}$ m. Finally, the depth of the bulk is highly variable and easily changed upwards from the millimetre scale and so $H$ is taken with the range of $1-10$ mm; estimates for the interfacial permeability, $\hat{\eta}$, are currently  unavailable. These dimensional parameter estimates generate the non-dimensional scales of Table~\ref{tab1}, which will guide the asymptotic and numerical investigations presented below.

\begin{table}[ht] \centering
\begin{tabular}{ccc}
\toprule
Parameter  &   Range  & Justification  \\
\midrule
$\hat{L}$ &  $8\times 10^{-3}$ m$- 2.2\times10^{-2}$ m &  See text \\
$\| \bm{\hat{D}_S}\|_\infty, \|\bm{\hat{D}_B} \|_\infty $ &   $10^{-10}$ m$^2$ s$^{-1}- 10^{-9}$ m$^2$s$^{-1}$   & Table~S8, \cite{grant2016orthogonal} \\
$1/\tau $ & $ 8.4\times 10^{-5}$ s$^{-1}-1.0\times10^{-3}$ s$^{-1}$ & Tables S8,~S9, \cite{grant2016orthogonal} \\
$H_\varepsilon$ &  $10^{-5}$ m$-  10^{-4}$ m & See text \\ 
$H$ &  $10^{-3}$m $- 10^{-2}$ m  &  See text \\
$\hat{\eta}$ & Unknown & $-$ \\
\bottomrule
\end{tabular}
\caption{Numerical scales of various  dimensional parameters and parameter groupings in SI units, based on patterning in synthetic pattern formation with {\it E. coli} bacterial colonies, using  physical scales motivated by the studies of  Grant et al.~\cite{grant2016orthogonal} and Boehm et al.~\cite{boehm2018}.}
  \label{tab0}
\end{table}

\begin{table}[ht] \centering
\begin{tabular}{cc}
\toprule 
Parameter &  Typical Value/Range \\
\midrule
$ \varepsilon = H_\varepsilon/\hat{L}$ & $4.5\times 10^{-4}-1.3\times 10^{-2} $ \\
$h = H/\hat{L}$ &  $0.045-1.3$ \\
$ \tilde{L}=L/\hat{L}$ & 1 \\
$\varepsilon_* = \varepsilon \| \bm{D_S}^{-1}\bm{J} \|^{1/2}_\infty$ &   $  1.1\times10^{-3}-  0.87$ \\
$h_* = h \| \bm{D_S}^{-1}\bm{J} \|^{1/2}_\infty$ &  $ 0.11-87$ \\
$\varepsilon_*^2/3$ &  $4.0 \times10^{-7}-0.3$  \\
$\|\bm{D_S}\|_\infty = \tau\| \bm{\hat{D}_S}\|_\infty/\hat{L}^2  $  &  $ 2\times 10^{-4}- 0.2$ \\ 
$\| \bm{D_B}\|_\infty \sim \|\bm{D_S}\|_\infty$ & $ 2\times 10^{-4}- 0.2$  \\ 
$\eta = \tau\hat{\eta}/\hat{L}$ & Unknown \\
\bottomrule
\end{tabular}

\caption{Numerical scales of various non-dimensional parameters and parameter groupings, motivated by the physical scales of synthetic pattern formation with {\it E. coli} bacterial colonies, in the studies of  Grant et al.~\cite{grant2016orthogonal} and Boehm et al.~\cite{boehm2018}. For matrices, the infinity (max) norm $\| \cdot \|_\infty$ is used, which is the modulus of the matrix component with largest magnitude. For the non-dimensional matrix Jacobian, this norm is taken to be of order unity as the timescale is non-dimensionalised relative to $\tau$, a representative timescale associated with  a fast reaction in the system. The non-dimensional lengthscale, $\tilde{L}$, is retained symbolically throughout the presentation to facilitate determining the impact of this scale, though it is unity for these scalings. The parameter scales $\varepsilon_*$ and $h_*$ as well as the range of $\varepsilon_*^2/3$ are presented as they will be   important in the asymptotic analyses below.  \label{tab1}}
\end{table}
\bk

\section{Linear Stability Analysis}
\label{GenLinAnalysis}

For a linear stability analysis of homogeneous equilibria of \eqref{eqS}-\eqref{coupling}, we require the steady states to this system, which arise from specifying
\begin{equation*}
\bm{f_S}(\bm{u_S^*}) = \bm{g}(\bm{u_S^*},\bm{u_B^*}) = \bm{u_S^*}-\bm{u_B^*}=\bm{0},\end{equation*}
so that the surface reactions determine the spatially-homogeneous steady state concentration in both regions, and our simple constitutive choice of $\bm{g}$ implies that these concentrations must be equal. We will focus exclusively on the case of an absence of reactions in the bulk, as motivated by the underlying inert agar layer in  synthetic pattern formation with {\it E. coli} bacterial experiments \cite{grant2016orthogonal,boehm2018} and which requires only a root of the surface kinetics for there to be a steady state.
 
We proceed by considering perturbations to this steady state of the form $$\bm{u_S} = \bm{u_S^*}+\sigma \bm{w_S}(x,y,t) , ~~~~~~~~\bm{u_B} = \bm{u_B^*}+\sigma \bm{w_B}(x,y,t),$$  where $|\sigma| \ll 1$, and in general the bulk and surface perturbations are $n$-dimensional functions,  where $n$ is the number of species. We substitute these perturbations into equations \eqref{eqS}-\eqref{coupling} to find, from equations \eqref{eqS}-\eqref{eqB}, that the perturbations will satisfy 
\begin{equation}\label{eqwS}
\frac{\partial \bm{w_S}}{\partial t} = \bm{D_S}\nabla^2 \bm{w_S}+\bm{J_{S}}\bm{w_S}, \quad x \in [0,\tilde{L}], \quad y \in [h,h+\varepsilon],
\end{equation}
\begin{equation}\label{eqwB}
\frac{\partial \bm{w_B}}{\partial t} = \bm{D_B}\nabla^2 \bm{w_B}, \quad x \in [0,\tilde{L}], \quad y \in [0,h],
\end{equation}
where the Jacobian, $\bm{J_{S}} = \partial \bm{f_S}/\partial \bm{u_S} \in \mathbb{R}^{n\times n}$, is evaluated at the steady state concentrations.  We also have the  coupling condition from equation~\eqref{coupling} given by,
\begin{equation}\label{couplingw}
\bm{D_S}\frac{\partial \bm{w_S}}{\partial y} = \eta(\bm{w_S}-\bm{w_B}), \quad -\bm{D_B}\frac{\partial \bm{w_B}}{\partial y} = \eta(\bm{w_B}-\bm{w_S}),  \text{ for } y=h.
\end{equation}

\subsection{Spatially homogeneous perturbations}\label{shp}

We now consider the appropriate generalisation of stability in the absence of transport, as is typically assumed in a Turing-type analysis. However, unless the reaction kinetics are the same in both domains,   which is not true in our setting,  then spatially homogeneous perturbations are not consistent with equations \eqref{eqwS}-\eqref{couplingw}. Such perturbations will not remain homogeneous under time evolution due to the coupling condition \eqref{couplingw}, except in the mathematically fine-tuned case where the homogeneous surface perturbation is along an eigenvector of $\bm{J_S}$ with eigenvalue $0$.
  
Previous studies of more complex systems (beyond those considered in textbook Turing models) also highlight that, when generalising the conditions that arise from the stability of the homogeneous steady state with respect to spatially homogeneous perturbations,  one must also consider a perturbation with respect to the zero mode(s) of the transport operator \cite{klika2018domain}. However, given the  assumption of completeness, i.e.~that separable solutions in $x$, $y$ and $t$ span the space of possible solutions, as generally used in linear stability theory,  the existence of zero modes of the transport operator also requires mathematical fine tuning. In particular, with $\nabla^2 \bm{u_S}^0=0$ for the zero mode of the transport operator acting in the surface layer, $\bm{u_S}^0$, one has
$$ 
\bm{u_S}^0 ={\bm A}\cos(k_qx)\cosh(k_q(y-(h+\varepsilon))), ~~~ k_q =q\pi/\tilde{L}, 
$$ for a general $\bm{A}$ and   $q$ a natural number  on enforcing the zero flux boundary conditions. There is a directly  analogous expression for the zero mode of the transport operator within the bulk region. However, after rearrangement, the interfacial condition at $y=h$ requires
\begin{equation*}
D_SD_B\sinh(k_q\varepsilon)\sinh(k_qh)=\eta \left(D_B \cosh(k_q\varepsilon)\sinh(k_qh)-D_S\sinh(k_q\varepsilon)\cosh(k_qh)
\right).
\end{equation*}
One possible solution occurs for $k_q=0$, which generates a spatially homogeneous mode that has already been considered above. Satisfying this equation  for other $k_q$  requires mathematical fine tuning as $k_q$ is already constrained to a set of zero measure and all other parameters are either geometrical or biophysical in origin.

Hence to summarise, in contrast to the textbook Turing case, unless the surface and bulk kinetics are the same, constraints on the parameters do not arise from the constraint of stability to homogeneous perturbations; instead this stability always holds, at least in the absence of a mathematical  fine tuning of  parameters and the possibility of such fine-tuning is neglected below.

\subsection{Spatially inhomogeneous perturbations}

To proceed, we \bk 
  assume a separable solution in $x$, $y$ and $t$ for the linearised system \eqref{eqwS}-\eqref{couplingw}.  With the usual assumption of a uniform temporal growth rate $\lambda$, the perturbation ansatz for a single mode of a separable solution is
\begin{equation}\label{expansions}
\bm{w_S} = e^{\lambda t} \bm{s}(y)\cos \left( k_qx\right), \, \bm{w_B} =  e^{\lambda t} \bm{b}(y)\cos \left( k_qx\right),
\end{equation}
 where $k_q  = q  \pi /\tilde{L} $ for $q$ a natural   number (including $0$). Assuming completeness of such a set of modes, then linear superposition entails that an arbitrary function may be expanded via a weighted linear sum of individual modes. Hence, the question of stability of a linear perturbation reduces to the same question for single modes, as in the standard textbook analysis \citep{Murray2003}, without the need for the modes to be orthogonal. Noting that homogeneous modes are not feasible, as shown above, we proceed to consider whether any heterogeneous modes exhibit instability ($\Re(\lambda)>0$). Furthermore we note that even in the absence of completeness, $\Re(\lambda)>0$ still provides a sufficient condition for instability, though it is not strictly necessary. We will see later that our conditions are not refuted by comparisons with numerics, and so we anticipate that the set of modes we construct is at least generic if not a complete basis. We note that for each $q$, there may be many distinct $\lambda$, and corresponding to each distinct pair of $(q,\lambda)$ we will have possibly different eigenfunctions $\bm{s},$ and $\bm{b}$. We will suppress this dependence in the following, but it is important to keep in mind that the following analysis applies for a given pair $(q,\lambda)$. As we are looking for modes which grow in time, leading to instability, we will  impose $\Re(\lambda)>0$ in the following. Substituting these expansions into \eqref{eqwS}-\eqref{eqwB} we find that a given mode satisfies,
\begin{equation}\label{eqwS1}
\lambda \bm{s} = \bm{D_S}(-k_q^2+\partial_y^2) \bm{s}+\bm{J_{S}}\bm{s}, \quad y \in [h,h+\varepsilon],
\end{equation}
\begin{equation}\label{eqwB1}
\lambda \bm{b} = \bm{D_B}(-k_q^2+\partial_y^2) \bm{b}, \quad y \in [0,h].
\end{equation}


After multiplying these equations by the inverse of the diffusion matrices and rearranging, we find 
\begin{equation}\label{eqwS2}
\bm{s}'' = (k_q^2\bm{I_n}+\bm{D_S}^{-1}(\lambda \bm{I_n}-\bm{J_{S}}))\bm{s}, \quad y \in [h,h+\varepsilon],
\end{equation}
\begin{equation}\label{eqwB2}
\bm{b}'' = (k_q^2\bm{I_n}+\lambda \bm{D_B}^{-1})\bm{b}, \quad y \in [0,h],
\end{equation}
where $'$ denotes the ordinary derivative with respect to $y$. These spatial functions  are required to satisfy the external boundary conditions $\bm{b}'(0)=\bm{s}'(h+\varepsilon)=\bm{0}$ and the coupling conditions which read,
\begin{equation}\label{couplingw2}
\bm{D_S}\bm{s'} = \eta(\bm{s_\textbf{q}}-\bm{b_\textbf{q}}), \quad \bm{D_B}\bm{b_\textbf{q}'} = \eta(\bm{s_\textbf{q}}-\bm{b_\textbf{q}}),  \text{ for } y=h.
\end{equation}

To find suitable $(\lambda,q)$ that solve the coupled problem \eqref{eqwS2}-\eqref{couplingw2}, we will make use of the matrix-valued function defined by $\cosh(\bm{M}) = (\exp(\bm{M})+\exp(\bm{-M}))/2$, for some matrix $\bm{M}$, as well as $\sinh(\bm{M}) = (\exp(\bm{M})-\exp(\bm{-M}))/2$. We recall the differentiation identity $\cosh(y\bm{M})' = \bm{M}\sinh(y\bm{M})$, which follows from this definition. We now seek to take the square-root of the matrices on the right-hand side of  equations \eqref{eqwS2} and \eqref{eqwB2} and thus define
\begin{equation}\bm{M_{S}}^2 = (k_q^2\bm{I_n}+\bm{D_S}^{-1}(\lambda\bm{I_n}-\bm{J_{S}})), \quad \text{and} \quad
\bm{M_{B}}^2= (k_q^2\bm{I_n}+\lambda\bm{D_B}^{-1}).
\end{equation}

We next consider solutions to equations \eqref{eqwS2} and \eqref{eqwB2} via hyperbolic matrix functions. As we will observe (e.g.~equation~\eqref{detcond} and the resulting dispersion relation), these matrices will always be in terms of functions that can be expressed in terms of even powers of $\bm{M_{B}}$ and $\bm{M_{S}}$, and thus functions of $\bm{M_{B}}^2$ and $\bm{M_{S}}^2$. This dependence on the squares of these matrices follows as if $f:\mathbb{C}\to\mathbb{C}$ is analytic, then a matrix-valued function can be defined via a power series in the matrix argument   \cite{higham2008functions}. 

The hyperbolic functions we will use are meromorphic with poles away from $0$, and hence the ambiguity  in defining the square root matrices, $\bm{M_{S}}$ and $\bm{M_{B}}$   does not play a role. Without loss of generality, we will consider the principal square roots of the matrices for definiteness, so that eigenvalues of $\bm{M_{B}}$ and $\bm{M_{S}}$ are the square roots with positive (or possibly zero) real parts of the eigenvalues of $\bm{M_{B}}^2$ and $\bm{M_{S}}^2$.

Proceeding, we then have the following solutions to equations \eqref{eqwS2} and \eqref{eqwB2} given by the hyperbolic matrix functions:
\begin{equation}\label{eigenfunctions}
\bm{s} = \cosh((y-h-\varepsilon)\bm{M_{S}})\bm{\alpha}, \quad \bm{{b}} = \cosh(y\bm{M_{B}})\bm{\beta},
\end{equation}
for some nonzero constant vectors $\bm{\alpha}, \bm{\beta}$. We note these functions satisfy the no-flux conditions at the top and bottom boundaries by construction. We now use the coupling conditions \eqref{couplingw} to determine a condition for nontrivial $\bm{\alpha}$ and $\bm{\beta}$. These read,
\begin{equation}\label{couplingwS}
-\bm{D_S}\bm{M_{S}}\sinh(\varepsilon\bm{M_{S}})\bm{\alpha} = \eta(\cosh(\varepsilon\bm{M_{S}})\bm{\alpha}-\cosh(h\bm{M_{B}})\bm{\beta}), 
\end{equation}
\begin{equation}\label{couplingwB}
    \bm{D_B}\bm{M_{B}}\sinh(h\bm{M_{B}})\bm{\beta} = \eta(\cosh(\varepsilon\bm{M_{S}})\bm{\alpha}-\cosh(h\bm{M_{B}})\bm{\beta}).
\end{equation}

We then have, writing equations \eqref{couplingwS} and \eqref{couplingwB} as a $2n\times 2n$ block matrix, the following condition for nontrivial solutions to this system:
\begin{equation}\label{detcond}
 \det{\begin{pmatrix} \eta\cosh(\varepsilon\bm{\bm{M_{S}}})+\bm{D_S}\bm{M_{S}}\sinh(\varepsilon\bm{M_{S}})  & -\eta \cosh(h\bm{M_{B}}) \\
\eta \cosh(\varepsilon \bm{M_{S}}) & -\eta\cosh(h\bm{M_{B}})-\bm{D_B}\bm{M_{B}}\sinh(h\bm{M_{B}})
\end{pmatrix}} = 0.
\end{equation}
As this condition involves transcendental functions of $\lambda$, we note that in general for a fixed spatial mode $q$, there will be infinitely many values of $\lambda$ for which equation~\eqref{detcond} is satisfied. Equivalently, $q$ only differentiates between eigenmodes in the $x$ direction, but cannot do so in $y$, and so these eigenmodes must be captured via multiplicity in $\lambda$.

While the condition given by equation~\eqref{detcond} is in principle computable, it is difficult to use to gain insight into Turing-like instabilities. Even simplifying the determinant condition is nontrivial, as the four blocks will not in general commute, so we now exploit the assumption of no reactions in the bulk to simplify this condition. We have that $\bm{M_{B}}^2$ is diagonal, and from our assumption that $\Re(\lambda)> 0$, we have that its eigenvalues have positive real part. Therefore, the elements of $\cosh(\bm{M_{B}})$ are given by the hyperbolic cosine of the diagonal elements of $\bm{M_{B}}$, and since these are all positive definite,  $\cosh(h\bm{M_{B}})$ is invertible.

Now we define the matrices $\bm{A} = \eta\cosh(\varepsilon\bm{M_{S}})+\bm{D_S}\bm{M_{S}}\sinh(\varepsilon\bm{M_{S}}) $, $\bm{B} = -\eta \cosh(h\bm{M_{B}})$, $\bm{C} = \eta \cosh(\varepsilon \bm{M_{S}})$, $\bm{D} = -\eta\cosh(h\bm{M_{B}})-\bm{D_B}\bm{M_{B}}\sinh(h\bm{M_{B}})$. By the above argument, we have that $\bm{B}$ is invertible. We then have that \eqref{detcond} can be written (by exchanging rows and using the Schur complement) as,
\begin{equation}\label{detcondsimp}
 \det{\begin{pmatrix} \bm{A} & \bm{B} \\
\bm{C} & \bm{D}
\end{pmatrix}} = (-1)^n\det(\bm{B})\det(\bm{C}-\bm{D}\bm{B}^{-1}\bm{A}) = 0.
\end{equation}
Noting that $\bm{D}\bm{B}^{-1} = \bm{I_n}+\bm{D_B}\bm{M_{B}}\tanh(h\bm{M_{B}})/\eta$, we have that 
\begin{align}
    \bm{C}-\bm{D}\bm{B}^{-1}\bm{A} =& \eta \cosh(\varepsilon \bm{M_{S}})\nonumber - \left( \bm{I_n}+\frac{1}{\eta}\bm{D_B}\bm{M_{B}}\tanh(h\bm{M_{B}})\right)(\eta\cosh(\varepsilon\bm{M_{S}})+\bm{D_S}\bm{M_{S}}\sinh(\varepsilon\bm{M_{S}}))\nonumber\\
    =& -\bm{D_B}\bm{M_{B}}\tanh(h\bm{M_{B}})\left(\cosh(\varepsilon \bm{M_{S}})+\frac{1}{\eta}\bm{D_S}\bm{M_{S}}\sinh(\varepsilon \bm{M_{S}})\right)\nonumber -\bm{D_S}\bm{M_{S}}\sinh(\varepsilon \bm{M_{S}}),\nonumber
\end{align}
so equation~\eqref{detcondsimp} is equivalent to,
\begin{equation}\label{detconfFull}
    \det\left(\bm{D_B}\bm{M_{B}}\tanh(h\bm{M_{B}})\left(\cosh(\varepsilon \bm{M_{S}})+\frac{1}{\eta}\bm{D_S}\bm{M_{S}}\sinh(\varepsilon \bm{M_{S}})\right)+\bm{D_S}\bm{M_{S}}\sinh(\varepsilon \bm{M_{S}})\right)=0.
\end{equation}
 We note that the Turing instability conditions for the surface system in isolation -- neglecting spatial structure in $y$ -- are precisely that the growth rates $\lambda$ computed from $\det(\bm{M_{S}})=0$   have negative real part for $k_0=0$, and  positive real part for some $k_q > 0$, and so this matrix encodes directly the classical case in this way. Furthermore, for  a fixed $q$, and with fixed model parameters, we expect that condition \eqref{detconfFull} admits infinitely many distinct values of $\lambda$. The intuition for this is that in the uncoupled case ($\eta=0$), the surface domain is a rectangle and, hence, the surface eigenfunctions $\bm{s}(y)$ are also cosines of different spatial eigenvalues, which can vary independently from $k_q$. However, we know of no method to compute analytical expressions for such spatial eigenvalues in the coupled case, and so instead use condition \eqref{detconfFull} to compute $\lambda$ directly, remaining aware of the inherent multiplicity.  To further understand the dispersion relation given by \eqref{detconfFull}, and how it relates to classical conditions for Turing instabilities, we now pursue several asymptotic reductions.



\section{Instability Conditions in Thin Surface Regimes}\label{Asymptotics}

In this section we compute instability conditions from equation~\eqref{detconfFull} for a variety of distinguished limits modelling a thin surface region, as motivated by synthetic patterning in bacterial populations. First, we mention even simpler reductions of the system, as a consistency check of our dispersion relation. We show that patterning is equivalent in the limit of decoupling the interaction of the surface and bulk regions, that is for sufficiently small $\eta \ll 1$. This is pursued in Appendix~\ref{App}, 
where  the classical Turing conditions are recovered as the surface system becomes isolated, as required.  In addition,  in Appendix~\ref{App}, we also demonstrate that no patterning can occur for classical Turing kinetics once all diffusion coefficients are equal in each of the regions, a direct analogue of the well known result that the classical Turing instability requires differential transport.

Noting that the full system is too rich to investigate in generality and that the non-dimensional surface depth parameters, $\varepsilon$ and $\varepsilon_*$  are small in Table~\ref{tab1} for the motivating example of synthetic pattern formation in {\it E. coli} colonies, we proceed below to studying pattern formation instabilities with thin surface asymptotics. In the experimental setting of Grant et al.~\cite{grant2016orthogonal}, the bacterial layer is always relatively thin, owing to transport constraints in the bacteria, though the agar layer can take different bulk heights. For this reason, after first introducing a thin surface limit of the dispersion relation \eqref{detcond} in Section~\ref{surfaceasy}, we consider subsequent limits of large or small bulk thickness, $h$, in Section~\ref{bulkasy}. We anticipate that the permeability of the interface, $\eta$, is large in these experiments but do not have quantitative estimates, and so also consider our asymptotics across varying values of this parameter. In Section~\ref{tsar}, we derive asymptotic results under a regular asymptotic assumption on $k$ and $\lambda$ (i.e.~that they remain comparable with non-asymptotic terms in the dispersion relation), and collect these results in Table~\ref{tablelims}. Finally in Section~\ref{ftsar}, we give an example of distinguished limits where this asymptotic assumption breaks down.
Throughout the following, we implicitly assume that the surface Jacobian, ${\bm J_S}$, has elements that are of the same order and thus of the order of $\|{\bm J_S}\|_\infty$, so that $\|{\bm J_S}{\bm A}\|_\infty$ is of the same scale as $\|{\bm J_S}\|_\infty \|{\bm A}\|_\infty$ for any matrix ${\bm A}$ considered.

\subsection{Thin Surface Limits $\left(\varepsilon_*^2/3 \ll 1\right)$}\label{surfaceasy}
  
Here, we consider an asymptotically thin-surface, requiring  $\varepsilon\|\bm{M_{S}}\|_\infty=H_{\varepsilon}\|\bm{M_{S}}\|_\infty/\tilde{L}\ll 1$. First note that  in the thin layer limit below, the surface Jacobian ${\bm J_S}$ only appears via 
\begin{eqnarray}\label{dsms} \bm{D_S}\bm{M_{S}}^2 =   k_q^2\bm{D_S}+ \lambda\bm{I_n}-\bm{J_{S}}.
\end{eqnarray}
In addition,  given patterning (i.e.~$\Re(\lambda)>0$),  the matrix  $ \bm{J_S}$ cannot be dominated by the terms $ \lambda  \bm{I_n}$ 
or $k_q^2\bm{D_S}$ within $ \bm{D_S}\bm{M_{S}}^2, $ since then the reaction kinetics are subleading in the requirements for patterning, which thus contain only terms associated with pure diffusion at leading order. However, pure diffusion cannot induce patterning, as demonstrated in Appendix \ref{AppB}. Thus we conclude that,  given patterning 
\begin{eqnarray}\label{cns} \mbox{max}(|\lambda|, k_q^2\|\bm{D_S} \| _\infty) \sim O(\| \bm{J_{S}}\| _\infty ),
\end{eqnarray} 
and also that
 $||\bm{M_S}||_\infty$ has an upper bound ( and in particular the $k_q^2$ term is in fact bounded).
Noting the boundedness of $\bm{M_S}$ we have that $\cosh(\varepsilon\bm{M_S})$ is invertible, as for $\varepsilon$ sufficiently small this matrix has a determinant which is asymptotically $1+\varepsilon^2 \textrm{trace}(\bm{M_S}^2)/2>0$. In addition, for sufficiently small $\varepsilon$, we have the  the   Taylor expansion 
\begin{eqnarray}\label{tanhexp}\tanh(\varepsilon {\bm M_S}) = \varepsilon {\bm M_S}\left(1+O\left(  \varepsilon^2 \|{\bm M_S}\|_\infty^2/3\right)\right), 
\end{eqnarray}
where $O( \varepsilon^2 \|{\bm M_S}\|_\infty^2/3 )$ means the same scale as $\varepsilon^2 \|{\bm M_S}\|_\infty^2/3 $, {\it or smaller}, as the surface thickness tends to zero. Thus by right muyltiplying \eqref{detcond} by $\cosh(\varepsilon\bm{M_S})^{-1}$ and Taylor expanding we obtain (to leading order) the relation
\begin{equation} \label{smalleps}
    \det\left(\bm{D_B}\bm{M_{B}}\tanh(h\bm{M_{B}})+\varepsilon\left(\frac{1}{\eta}\bm{D_B}\bm{M_{B}}\tanh(h\bm{M_{B}})+\bm{I_n}\right)\bm{D_S}\bm{M_{S}}^2\right)=0, 
\end{equation} 
providing $\varepsilon^2\|\bm{M_{S}}\|^2_\infty/3\ll 1$ (ensuring the invertibility of $\cosh(\varepsilon \bm{M_S})$ and the validity of the Taylor expansion above).
Furthermore, noting that  
$\varepsilon_* = \varepsilon \| \bm{D_{S}}^{-1}\bm{J}_S\|_\infty^{1/2}$ together with the relations  (\ref{cns}), 
which give the maximum scale of $k_q^2$ and show that  
$|\lambda| \| \bm{D_{S}}^{-1}\|  \sim O(\| \bm{D_{S}}^{-1}\|_\infty \|\bm{J}_S\|_\infty )\sim O(\| \bm{D_{S}}^{-1}\bm{J}_S\|_\infty ),$ 
we have 
\begin{eqnarray}  
\varepsilon^2\|\bm{M_{S}}\|^2_\infty \sim \mbox{max}\left(\varepsilon^2  k_q^2, \varepsilon^2\| \bm{D_{S}}^{-1}\bm{J_S} \| _\infty   \right)
\sim 
\mbox{max}\left(\varepsilon^2  \frac{\| {\bm J_{S}} \| _\infty }{\| \bm{D_{S}} \| _\infty }, \varepsilon^2 \| {\bm D_{S}}^{-1} \| _\infty   \| {\bm J_{S}} \| _\infty    \right)\sim\varepsilon^2 \| \bm{D_{S}}^{-1}\bm{J}_S\|_\infty  =\varepsilon^2_* ,   \label{reliq1} 
\end{eqnarray} 
\bk using $ \| {\bm D_{S}}^{-1} \| _\infty\geq 1/ \| {\bm D_{S}}  \| _\infty.$
The latter inequality is immediate in the two species case on writing ${\bm{D_S}} = \mbox{diag}(a,a\xi)$ with $\xi \leq 1,$ as then $\| {\bm D_{S}}^{-1} \| _\infty = 1/(a\xi) \geq 1/a =1/\| {\bm D_{S}}  \| _\infty,$ with a trivial generalisation to higher number of species. Hence, for conditions associated with patterning, the relative error in the leading order thin surface approximation arising from equation (\ref{tanhexp}) is $\varepsilon_*^2/3 $  and thus we require  $\varepsilon_*^2/3 \ll 1.$
Despite the very large range of potential parameters in Table~\ref{tab1}, the scales for synthetic patterning in bacterial colonies are consistent with this bound.

\subsection{Consideration of bulk depth $h$}\label{bulkasy}

Noting $\bm{D_S} \approx \bm{D_B}$ at least for the parameter estimates of Tables \ref{tab0}, \ref{tab1}, and also relations (\ref{cns}),  (\ref{reliq1}) we also have   
\begin{eqnarray}\label{cns1} \|\bm{M_{B}}^2\| _\infty =  \| k_q^2     \bm{I_n}+\lambda\bm{D_B}^{-1}\| _\infty \sim O( \|\bm{D_S}^{-1} \bm{J_{S}}\|_\infty), ~~~~~~~~~~  \|\bm{D_{B}}\bm{M_{B}}^2\| _\infty  =  \| k_q^2     \bm{D_B}+\lambda\bm{I_n}\| _\infty  \sim O( \|  \bm{J_{S}}\|_\infty) .
\end{eqnarray}
Hence an appropriate scale for the largest component of $h\bm{M_B}$ is 
$h_* = h \| \bm{D_S}^{-1}\bm{J_S} \|^{1/2}_\infty$, which ranges from small to large in Table~\ref{tab1} and thus we proceed to consider simplifications of the expression $ \bm{M_B}\tanh(h \bm{M_{B}})$ within the instability condition (\ref{detconfFull}) for small and large values of $h_*$.  
For the small $h_*$ limit a Taylor series expansion immediately  gives 
$\bm{M_{B}}\tanh(h \bm{M_{B}})\sim  h \bm{M_{B}}^2$, with relative corrections on the scale of $h_*^2/3$ and we also have  
\begin{eqnarray}
\label{hs1}
\|\bm{M_{B}}\tanh(h \bm{M_{B}})\|_\infty\sim  h \|\bm{M_{B}}^2\|_\infty   \sim h_* \|\bm{M_{B}} \|_\infty  ~~~~ \mbox{for}~~ h_*\ll1  .
\end{eqnarray}
 
For large $h_*$ simplifications, first note  that $\bm{M_B}^2$ is diagonal, with diagonal components that have positive real parts since  $\Re(\lambda)>0$ as we require instability. 
Furthermore, similar to the synthetic patterning explored in experimental studies \cite{grant2016orthogonal,boehm2018}, we are interested in lateral patterning (in the $x$-direction of Fig. \ref{Schematic}), thus, we take $k_q^2>0$ and enforce $k_q^2\geq \pi/\tilde{L}$ by wavemode selection, which bounds the real part of $\bm{M_B}^2$ away from zero.

For $z\in\mathbb{C}$ with $\Re(z)\neq 0$, we have the limit
$$ z\tanh(z) \rightarrow \mbox{Sign}(\Re(z))z \mbox{ ~~~as~~~ } |z|\rightarrow \infty   ,$$ 
as may be deduced by writing $z$ in terms of its real and imaginary parts, with subsequent use of the properties of trigonometric and hyperbolic functions. 
In addition we have, without loss of generality, defined   $\bm{M_B}$ by the diagonal matrix with  {\it positive} semi-definite real part for the square root of the diagonals of $\bm{M_B}^2$, 
and in fact  no such square root has zero real part since the diagonals of $\bm{M_B}^2$ have positive real part. Consequently, at leading order we have in the large $h_*$ limit that 
$h\bm{M_B}\tanh(h \bm{M_{B}}) \rightarrow h\bm{M_B}$  and thus $\bm{M_B}\tanh(h \bm{M_{B}}) \rightarrow \bm{M_B}$, with 
\begin{eqnarray}\label{hs2} 
\|\bm{M_B}\tanh(h \bm{M_{B}})\|_\infty \sim \|\bm{M_B}\|_\infty  ~~~~ \mbox{for}~~ h_*\gg 1. 
\end{eqnarray} 
Finally, with this definition of $\bm{M_B}$, which is diagonal with terms whose real parts are bound away from zero, we also have that the diagonal elements, and hence the matrix norm,  do not blow up on taking the hyperbolic tangent (all of its singularities lie on the imaginary axis) and thus $ \| \tanh(h \bm{M_{B}})\|_\infty  \sim O(1) $ for $h_* \sim O(1)$. This may be summarised together with equations \eqref{hs1} and  \eqref{hs2} via    
\begin{eqnarray}\label{hs3} 
\|\bm{M_B}\tanh(h \bm{M_{B}})\|_\infty \sim \mbox{min}(h_*\|\bm{M_B}\|_\infty ,  \|\bm{M_B}\|_\infty) =  \mbox{min}(h_*,1) \|\bm{M_B}\|_\infty.
\end{eqnarray}

We are now in a position to consider the small $\varepsilon_*$, thin surface,  limit of the instability condition given by equation (\ref{detconfFull}), considering the full range of values of $h_*$, which is a measure of the non-dimensional depth of the bulk relative to the patterning lengthscale. 
We also consider the case $h_* \sim O(\varepsilon_*)$ for relative completeness,  even though Tables \ref{tab0},  \ref{tab1} highlight  that $h_*\gg \varepsilon_*$   is  anticipated for experiments with synthetic pattern formation within  bacterial populations.
 
\subsection{Thin surface asymptotic regimes with $k_q^2 \|\bm{D_S}\| _\infty,~|\lambda| \sim $~ord$(\|\bm{J_S}\| _\infty)$}\label{tsar}

An example of patterning when $k_q^2 \|\bm{D_S}\| _\infty,~|\lambda| \ll  \mbox{ord}(\|\bm{J_S}\| _\infty) $ is given  in the next subsection, but here we consider thin surface asymptotics with $\varepsilon_*^2/3 \ll 1$ on fixing $k_q^2 \|\bm{D_S}\| _\infty,~|\lambda| \sim $~ord$(\|\bm{J_S}\| _\infty),$
where ord$(\|\bm{J_S}\| _\infty)$ is defined to mean  both $O(\|\bm{J_S}\| _\infty)$ and not $o(\|\bm{J_S}\| _\infty).$ Hence we are considering pattern formation that occurs on the timescales of the kinetics with a lengthscale associated with the timescale of the kinetics and the (largest) diffusion scale and, as previously noted,  this simplifies the instability condition \eqref{detconfFull} at leading order to 
\begin{equation}\label{smalleps1}
    \det\left(\bm{D_B}\bm{M_{B}}\tanh(h\bm{M_{B}})+\varepsilon\left(\frac{1}{\eta}\bm{D_B}\bm{M_{B}}\tanh(h\bm{M_{B}})+\bm{I_n}\right)\bm{D_S}\bm{M_{S}}^2\right)=0, 
\end{equation}
where the scale of the non-dimensional permeability $\eta$ is unknown, and the possible values of  $h_*=h \|\bm{D_S}^{-1}\bm{J_S}\|^{{ 1/2}}_\infty\sim h \|\bm{M_{B}}\|_\infty \ $
are wide-ranging.  Hence, there are several nontrivial distinguished limits, which we proceed to document. Where possible, we will also relate these limits to the isolated surface case, where $\lambda$ is determined by the dispersion relation $\det(\bm{M_S^2})=\det(\lambda\bm{I_n}+k_q^2\bm{D_S}-\bm{J_{S}})=0$, in order to understand the impact of the bulk on the classical single-domain situation.

\vspace{6pt}\noindent{\bf Case I  $h_* \ll \varepsilon_* \ll 1$:} Noting that  $h_* \ll \varepsilon_*$ is equivalent to $h \ll \varepsilon$ by definition, in this limit equation  \eqref{smalleps1} reduces to,
\begin{equation}\label{smallepshleps}
    \det\left(\left(\frac{1}{\eta}\bm{D_B}\bm{M_{B}}\tanh(h\bm{M_{B}})+\bm{I_n}\right)\bm{D_S}\bm{M_{S}}^2\right)=\det\left(\frac{1}{\eta}\bm{D_B}\bm{M_{B}}\tanh(h\bm{M_{B}})+\bm{I_n}\right)\det\left(\bm{D_S} \right)\det\left( \bm{M_{S}}^2\right)=0.
\end{equation}
However the determinant with the hyperbolic tangent term cannot generate a root with $\Re(\lambda)>0$ and thus patterning. In particular, 
in Appendix \ref{AppA2}, following equation (\ref{ztz}),  it is shown that when $\Re (z^2)>0$ one also has $\Re (z\tanh(z))>0.$ With $\Re(\lambda)>0$ for patterning, let $z^2=h^2(k_q^2+\lambda /d_B)$, where $d_B$ is a bulk diffusion coefficient. Thus $z^2$ is an eigenvalue of $h^2\bm{M_{B}}^2$, and all eigenvalues of this matrix are of this form. Furthermore, we have $\Re(z^2)>0$ where $z$ is an eigenvalue of $h\bm{M_{B}}$ and satisfies $\Re (z\tanh(z))>0.$ However for the hyperbolic tangent term in equation (\ref{smallepshleps}) to generate a root, at least one  eigenvalue of $h\bm{M_{B}}$, that is one such $z$, must satisfy  $\Re (z\tanh(z))<0,  $ a contradiction,    thus showing there are no roots from the determinant involving the hyperbolic tangent.   Hence,  noting $ \bm{D_S}  $ is positive definite,
the only roots are those of the   isolated Turing modes, independent of $\eta$,  and  determined purely from $\det\left(\bm{M_{S}}^2\right)=0$.

\vspace{6pt}\noindent{\bf Case II  $\varepsilon_* \ll 1$, $h_*/ \varepsilon_*=h/\varepsilon = \hat{h} \sim \mathrm{ord}(1)$:} This limit corresponds to the entire domain being thin with respect to the lengthscale in the $x$ direction ($\tilde{L}$). In this case we have,
\begin{equation}\label{smallepsh}
    \det\left(\hat{h}\bm{D_B}\bm{M_{B}}^2+\left(\frac{\varepsilon\hat{h}}{\eta}\bm{D_B}\bm{M_{B}}^2+\bm{I_n}\right)\bm{D_S}\bm{M_{S}}^2\right)=0.
\end{equation}
Equation \eqref{smallepsh} is a slight modification of the isolated surface Turing conditions in 1-D given by $\det(\bm{M_{S}}^2)=0$, and can similarly be written as an $n$th order polynomial in $\lambda$. Further,  if 
$\eta \ll \varepsilon \| \bm{D_B} \bm{M_{B}}^2\|_\infty \sim \mbox{ord}(\varepsilon \| \bm{J_S} \|_\infty)$, then the conditions for instability are precisely those for an isolated surface. Similarly, if $\eta = \mbox{ord}(\varepsilon\| \bm{D_B} \bm{M_{B}}^2\|_\infty)\sim \mbox{ord}(\varepsilon\| \bm{J_S} \|_\infty)$, then we are left with a `quadratic' dispersion relation, which does not simplify from the form given in \eqref{smallepsh} (`quadratic' meaning this dispersion relation will give a polynomial of order $2n$ for $\lambda$, compared to the standard $n$th order polynomial). In general such a relation could lead to quite different values of $\lambda$ from the isolated case, though we will not analyse it further here.
If $\eta \gg \varepsilon\| \bm{D_B} \bm{M_{B}}^2\|_\infty\sim \mbox{ord}(\varepsilon\| \bm{J_S} \|_\infty) $, we then have the instability condition,
\begin{equation}\label{smallepshleta}
    \det\left(\lambda(1+\hat{h})\bm{I_n}+k_q^2(\hat{h}\bm{D_B}+\bm{D_S})-\bm{J_{S}}\right) = 0,
\end{equation}
which can be seen as a homogenisation, or averaging, of the bulk and surface layers. Such an averaged dispersion relation has the potential to increase the ability of the system to pattern compared to the isolated case by, e.g., introducing, or increasing, the differential diffusion between species. 

In some other (experimentally relevant) cases this averaged system will decrease the ability of the system to pattern compared to the isolated case. For instance, the necessary differential diffusion for Turing patterning may be due to, for example, substrate binding {\gr \cite{korvasova}}   that is only present in the surface system. In an inert bulk region, there are fewer physical scenarios where differential diffusion is likely as most biological proteins are roughly the same size. In such a case, we have that $\bm{D_B}=c_B\bm{I}$, so that \eqref{smallepshleta} can be rearranged to given,
\begin{equation}\label{smallepshletaDB=I}
    \det\left(\left(\lambda(1+\hat{h})+k_q^2\hat{h}c_B\right)\bm{I_n}+k_q^2\bm{D_S}-\bm{J_{S}}\right) = 0,
\end{equation}
which we can see as a shrinking and shifting to the left a root $\lambda$ coming from the isolated case. Effectively then, such a scenario leads to a smaller instability region in parameter space,  subject to the wavemode selection constraint that $k_q=q\pi/\tilde{L}$, for a natural number $q$.

Another plausibly relevant case of equation~\eqref{smallepshleta} is if $\bm{D_S}=\bm{D_B}$, i.e.~the surface and bulk diffusivities are the same.  Here, the dispersion relation 
is that of the classic case except $\lambda$ and $k_q^2$ are both scaled by $(1+\hat{h})$. Hence the allowed values of $(\lambda,k_q^2)$ are those of the classic case divided by $(1+\hat{h})$, which shrinks the range of the allowed patterning wavenumbers relative to the classic case and thus leads to a smaller Turing space compared to the isolated surface system,  though again subject to the wavemode selection constraint.

\vspace{6pt}\noindent
{\bf Case III   $\varepsilon_* \ll h_* $ : }  This case proceeds similarly regardless of whether $h_* \ll 1, ~h_*\sim\mbox{ord}(1)$ or $h_*\gg 1$.  
Noting  
$\bm{D_S} \approx \bm{D_B}$,  
$ \|\bm{D_S}\bm{M_{S}}^2\|_\infty \sim   \|  \bm{J_S} \|_\infty   , $ from equations (\ref{dsms}) and  (\ref{cns}), 
$\|\bm{M_{B}} \| _\infty  \sim O( \|\bm{D_S}^{-1} \bm{J_{S}}\|^{1/2}_\infty) $ 
by square rooting the first of relations (\ref{cns1}), 
 and equation (\ref{hs3}), that is  $\|\bm{M_B}\tanh(h \bm{M_{B}})\|_\infty \sim    \mbox{min}(h_*,1) \|\bm{M_B}\|_\infty \sim    \mbox{min}(h_*,1) \|\bm{D_S}^{-1} \bm{J_{S}}\|^{1/2}_\infty $, 
we have  
$$\frac{\varepsilon\| \bm{D_S} \bm{M_{S}}^2\|_\infty}{ \| \bm{D_B} \bm{M_{B}}\tanh(h \bm{M_{B}})\|_\infty } \sim 
\frac{\varepsilon \|   \bm{J_S} \|_\infty}{\|\bm{D_S}\|_\infty\|\bm{D_S}^{-1} \bm{J_{S}}\|^{1/2}_\infty \mbox{min}(h_*,1)}  \sim 
\frac{\varepsilon \|  \bm{D_S}^{-1} \bm{J_S} \|_\infty}{\|  \bm{D_S}^{-1}   \|_\infty\|\bm{D_S}\|_\infty\|\bm{D_S}^{-1} \bm{J_{S}}\|^{1/2}_\infty \mbox{min}(h_*,1)}. 
$$
Noting $ \| {\bm D_{S}}^{-1} \| _\infty \| {\bm D_{S}}  \| _\infty \geq 1,$ as deduced just below equation (\ref{reliq1}), and $ \varepsilon_* = \varepsilon\|  \bm{D_S}^{-1} \bm{J_S} \|_\infty^{1/2} $ we thus have 
$$
\frac{\varepsilon\| \bm{D_S} \bm{M_{S}}^2\|_\infty}{ \| \bm{D_B} \bm{M_{B}}\tanh(h \bm{M_{B}})\|_\infty } 
\lesssim \frac{\varepsilon_*}{  \mbox{min}(h_*,1)} 
\sim \mbox{max}\left(\varepsilon_*,\frac{\varepsilon_*}{h_*}\right)  \ll 1.
$$
Hence the final term from relation \eqref{smalleps}, that is $\varepsilon\bm{D_S}\bm{M_{S}}^2$, may always  be dropped relative to the first term, that is $ \bm{D_B}\bm{M_{B}}\tanh(h\bm{M_{B}})$. 
This reveals that the instability condition simplifies to  
\begin{equation}\label{smallepsord1h}
    \det\left(\bm{D_B}\bm{M_{B}}\tanh(h\bm{M_{B}})\left(\frac{\varepsilon }{\eta}\bm{D_S}\bm{M_{S}}^2+\bm{I_n}\right)\right) = \det(\bm{D_B}\bm{M_{B}}\tanh(h\bm{M_{B}}))\det\left(\frac{\varepsilon }{\eta}\bm{D_S}\bm{M_{S}}^2+\bm{I_n}\right)  = 0.
\end{equation}
The hyperbolic tangent  does not contribute to instability, by analogous reasoning to Case I. Thus, there is no  instability unless $\eta$ is concomitantly small alongside $\varepsilon\| \bm{D_S} \bm{M_{S}}^2\|_\infty\sim \varepsilon\| \bm{J_S}  \|_\infty $. Writing out $\bm{M_{S}}^2$, we see that the impact of the bulk on the surface system is simply to shift the eigenvalues to the left in the complex plane by the quantity $\eta/\varepsilon  $, and hence the Turing space for this system is strictly smaller than the Turing space for an isolated one-dimensional system with surface kinetics.

Collecting all of these various limits together in Table~\ref{tablelims}, we can see a pattern emerging. As $\eta$ or $h_*$ is increased, we observe a trend of moving from the isolated surface system to a reduced, or average system, and eventually, for $h_* / \varepsilon_* \sim h/\varepsilon \gg 1$  and $\eta \gg \varepsilon\| \bm{J_S}  \|_\infty $, to no patterning being permitted. While it is not true in general that the Averaged case or the Quadratic case correspond to a reduced ability for a system to pattern, we anticipate that this is the case for most standard Turing systems, and hence there is a broadly monotonic decrease on the ability of a system to pattern as the bulk becomes larger or the boundary more permeable. This has concomitant implications for prospective multilayered Turing systems, such as the experimental studies involving bacterial patterning which motivate this study.

\begin{table}[ht]
\centering
\begin{tabular}{cccc} 
\toprule
& Case I.  &  Case II. & Case III. \\
& ($h_* \ll \varepsilon_* \ll 1$) & ($\varepsilon_* \ll 1$, $\varepsilon_*/h_*=\varepsilon/h \sim \mbox{ord}(1)$) & ($\varepsilon_* \ll h_*$) \\
\midrule
$ \eta \ll  \varepsilon \| \bm{J_S} \|_\infty $& Isolated  & Isolated  &Isolated     \\ 
$\eta \sim \mbox{ord}\ ( \varepsilon \| \bm{J_S} \|_\infty)$ &Isolated &Quadratic condition, Eqn (\ref{smallepsh})& Reduced instability     \\
$ \eta \gg  \varepsilon \| \bm{J_S} \|_\infty$&Isolated &Averaged condition, Eqn (\ref{smallepshleta}) &No instabilities \\ 
\bottomrule
\end{tabular}
\caption{Thin-surface limits obtained in different asymptotic regimes  given 
$k_q^2 \|\bm{D_S}\| _\infty,~|\lambda| \sim $~ord($\| \bm{J_S}  \|_\infty $).  Note that moving left to right corresponds to an increasing size of $h_* $, and moving top to bottom corresponds to increasing scales of $\eta$. No instabilities:  $\det(\bm{D_{B}}\bm{M_{B}}\tanh(h\bm{M_{B}}))=0$; Isolated (1-D) surface: $\det(\bm{M_S}^2)=0$; Quadratic $\lambda$: $\det(\hat{h}\bm{D_B}\bm{M_{B}}^2+(\hat{h}c\bm{D_B}\bm{M_{B}}^2+\bm{I_n})\bm{D_S}\bm{M_{S}}^2)=0$,   $c =  \varepsilon/\eta \sim  \mbox{ord}(\| \bm{J_S} \|_\infty^{-1})$;  Reduced instability: $\det(c\bm{D_S}\bm{M_{S}}^2+\bm{I_n})=0$, $c = \varepsilon/\eta \sim \mbox{ord}(\| \bm{J_S} \|_\infty^{-1})$; Averaged condition: equation~\eqref{smallepshleta}. }\label{tablelims}
\end{table}

\subsection{Further thin surface asymptotic regimes with $ |\lambda|,   k_q^2 \|\bm{D_S}\| _\infty\sim   \mbox{ord}\left(\varepsilon^{1/2} \|\bm{J_S}\| _\infty\right)$} \label{ftsar}

There exist nontrivial asymptotic limits which are not described by $|\lambda|, k^2_q \|\bm{D_S}\| _\infty \sim$~ord$(\|\bm{J_S}\| _\infty)$, which can lead to instabilities not captured in Table~\ref{tablelims}, as we now show. 
In particular, with   $|\lambda|, k_q^2\|\bm{D_S}\| _\infty \sim \mbox{ord}(\varepsilon^{1/2}\|\bm{J_S}\| _\infty)$ and $\bm{D_B}\approx \bm{D_S}$ we then have $\varepsilon \bm{D_S}\bm{M_S}^2 \sim -\varepsilon \bm{J_S} + \mbox{ord}(\varepsilon^{3/2}\|\bm{J_S}\| _\infty),$ $\|\bm{D_B}\bm{M_B}^2\|_\infty \sim \mbox{ord}( |\lambda|,   k_q^2 \|\bm{D_B}\| _\infty)\sim  \mbox{ord}(\varepsilon^{1/2}\|\bm{J_S}\| _\infty)$ and finally $\|h\bm{M_B}\|_\infty \sim  \mbox{ord}(h\varepsilon^{1/4}\|\bm{D_S}^{-1} \bm{J_S}\| _\infty^{1/2})
  \sim \mbox{ord}(h \varepsilon_*/\varepsilon^{3/4})$. Hence from equation~\eqref{smalleps}, after expanding $\tanh(z) \sim z\left(1+ \mbox{ord}\left(z^2/3\right)\right)$ for small $z$, we have at leading order 
\begin{equation}\label{smallepslambdakq}
    \det\left(h\bm{D_B}\bm{M_{B}}^2+\frac{\varepsilon h}{\eta}\bm{D_B}\bm{M_{B}}^2\bm{D_S}\bm{M_{S}}^2+ \varepsilon\bm{D_S}\bm{M_{S}}^2\right)=0,
\end{equation}
with relative corrections  of   $ h ^2\varepsilon_*^{2} /(3\varepsilon^{3/2}) $
which is required to be much less than unity. 
Further,  noting we have already assumed $h  \sim \mbox{ord}(\varepsilon^{1/2})$,  we thus additionally require  $ \varepsilon_*^{2}/(3\varepsilon^{1/2}) \ll 1$ for  equation (\ref{smallepslambdakq}) to hold.  For the range of parameters detailed in Table~\ref{tab1}, we have  $ \varepsilon_*^{2}/(3\varepsilon^{1/2}) \in[1.8\times 10^{-5},2.3]$, and thus we have equation (\ref{smallepslambdakq}) is typically valid  for parameters  associated with synthetic patterning in bacterial colonies, but not always.

We proceed by noting that the first and third terms of equation (\ref{smallepslambdakq})  are $\mbox{ord}(\varepsilon\|\bm{J_S}\| _\infty)$ and the second is $\mbox{ord}\left(\varepsilon^2\|\bm{J_S}\| _\infty^2/\eta\right)$. 
 Writing  
$$
\lambda=\varepsilon^{1/2}   \mu, ~~~~~~~ h=\varepsilon^{1/2}\hat{h}, ~~~~~~~ k_q^2 \bm{D_B}= \varepsilon^{1/2}  K_q^2\bm{D_B} ~~~~~~~ \mbox{with} ~~~~|\mu|,   K_q^2\|\bm{D_B}\|_\infty  \sim~\mbox{ord}(\|\bm{J_S}\| _\infty), ~~~~  \hat{h} \sim~\mbox{ord}(1).$$
We then have $h\bm{D_B}\bm{M_B}^2 = \varepsilon  \hat{h}( K_q^2\bm{D_B}+\mu\bm{I_n}),$ and can factor   an $\varepsilon  $ from equation~\eqref{smallepslambdakq} to obtain,
\begin{equation}\label{smallepslambdakq2}
    \det\left(\hat{h}(K_q^2\bm{D_B}+\mu\bm{I_n})-\frac{\varepsilon  \hat{h}}{\eta}\left (K_q^2\bm{D_B}+\mu\bm{I_n}\right )\bm{J_{S}}- \bm{J_S}\right)=0,
\end{equation}
which, in general, can admit {\it nontrivial instabilities due to the coupling of the surface and the bulk.} In particular, when $\eta \gg \varepsilon \|\bm{J_S}\| _\infty$, so that the second term is no longer retained in the leading order, we find that the growth rates $\mu$ are given as the eigenvalues of $\bm{J_S}/\hat{h}-K_q^2\bm{D_B}$.

This matrix resembles the classical isolated-surface case except with a scaling of the kinetics by $\hat{h}$ and the appearance of the bulk diffusion parameters, rather than those in the surface. Hence, we can use  usual methods (e.g.~the Routh-Hurwitz criterion) to determine parameters that lead to instability in this case, noting that any values of $\lambda$ associated with instability  will be of modulus  $\mbox{ord}(\varepsilon^{1/2})$, and hence will be associated with slow growing modes.  Additionally, we anticipate that such modes will also exhibit small amplitude patterns, as is typical due to center-manifold reduction near Turing-type bifurcations \cite{cross1993pattern}, and hence may not be visible against experimental noise.
While other distinguished limits may exist which do not fall into the classifications given in Table~\ref{tablelims}, for brevity we do not pursue a systematic classification of these here. In the next section we will show that almost all numerically computed dispersion relations given by condition \eqref{detconfFull} fall within the asymptotics given in Table~\ref{tablelims},
with the exception of the case given in equation~\eqref{smallepslambdakq2} which was found numerically first, and subsequently  motivated the above  scaling. 


\section{Numerical Exploration of Example Systems}\label{Numerics}

As an example of these dynamics we consider the Schnakenberg kinetics for surface reactants $\bm{u_S} = (u_S, v_S)$ given by $$ \bm{f_S}(u_S, v_S) = \left(a-u_S+u_S^2v_S,b-u_S^2v_S\right)$$ with $a \geq 0$, $b > 0$. The spatially homogeneous steady state is given by $\bm{u_S^*} = \bm{u_B^*} = \left(a+b, b/(a+b)^2\right)$. Unless otherwise stated, we will assume equal diffusion coefficients between the surface and the bulk given by the diagonal matrices $\bm{D_S}=\bm{D_B}=\text{diag}(d_u, d_v)$. Without bulk reactions, and given linear interfacial conditions as summarised by equation (\ref{coupling}) with the relations (\ref{geqn}) and (\ref{sceqn}), we can immediately apply condition \eqref{detconfFull} to determine whether, or not, we expect a solution to pattern, and then compare these predictions with numerical simulations of the full nonlinear system.

Numerically computing $\lambda$ from condition \eqref{detconfFull} is substantially more involved than typical Turing-type analyses (e.g.~for polynomial dispersion relations \cite{Murray2003}) due to the transcendental nature of this determinant condition. In particular, we expect that for any given wavemode in the $x$ direction given by $k_q=q\pi/\tilde{L}$, for a natural number $q$, we have infinitely many distinct values of $\lambda$. These essentially correspond to the wavemodes in the $y$ direction which we have found only implicitly in our construction of the dispersion relation. So to determine if, for a given set of parameters, condition \eqref{detconfFull} admits a value of $\lambda$ with $\Re(\lambda)>0$ we resort to numerical heuristics. While fast general-purpose methods exist for rootfinding of polynomials over the complex numbers \cite{verschelde1999algorithm}, we are unaware of similar methods for more complicated functions. In lieu of this, we developed a set of numerical heuristics to accurately determine whether or not a value of $\lambda$ with $\Re(\lambda)>0$ exists, and tested this against full numerical simulations. We make use of the Matlab function \texttt{PatternSearch} as well as a deflation algorithm based on Muller's method to find many candidate roots with positive real part \cite{muller1956method, conte2017elementary}, and then discard any which are spurious. Throughout this section, we denote the largest such root by $\max(\Re(\lambda))$, noting that even in the classical case this maximum is needed as there are generically $n$ distinct values of $\lambda$.

\begin{figure} \centering
  \begin{tabular}{cc}
    \subfloat[Case I: $h=10^{-3}, \varepsilon=10^{-2}$] {\includegraphics[width=0.45\textwidth]{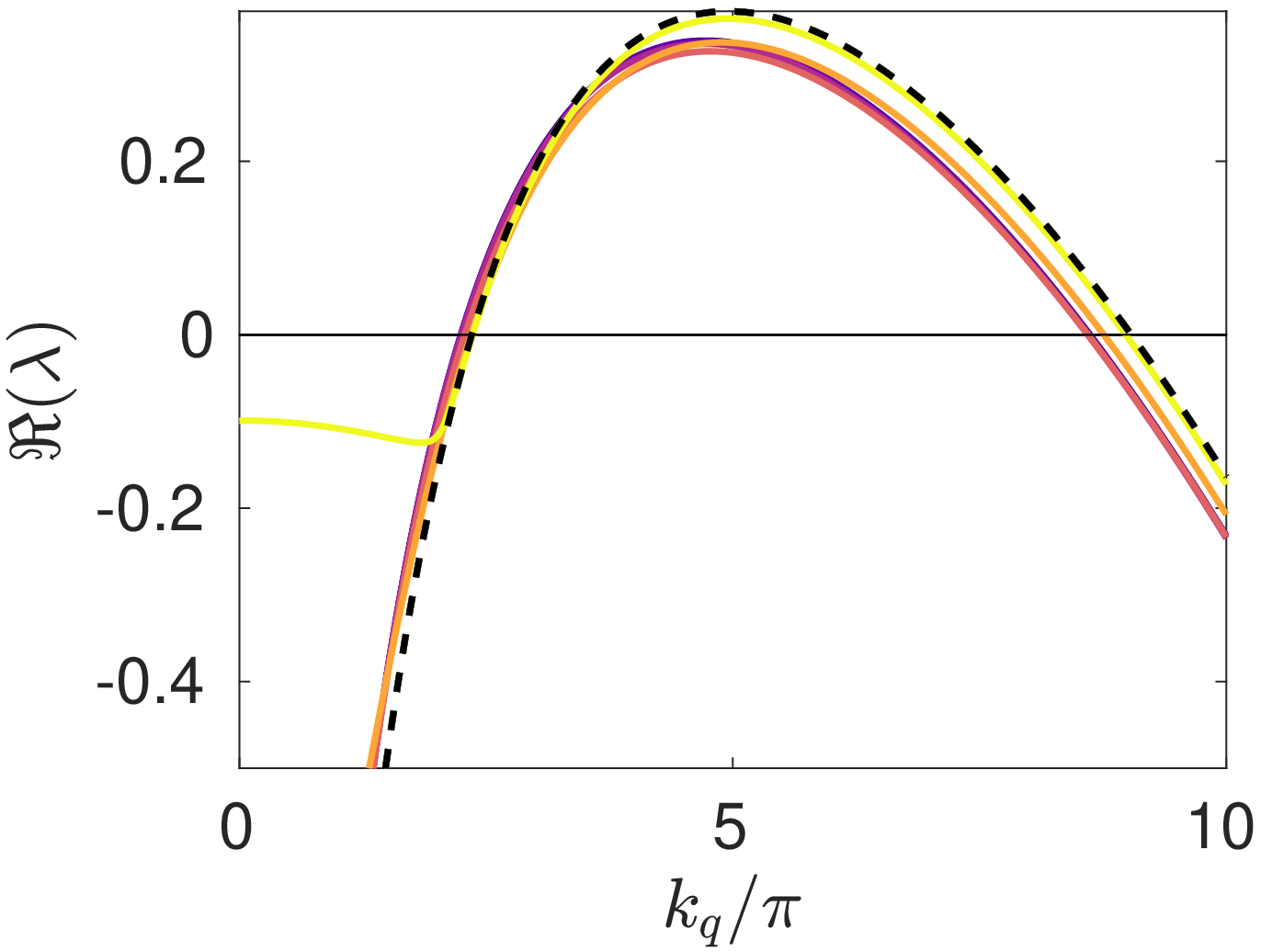}} &
    \subfloat[Case II: $h=10^{-2}, \varepsilon=10^{-2}$] {\includegraphics[width=0.45\textwidth]{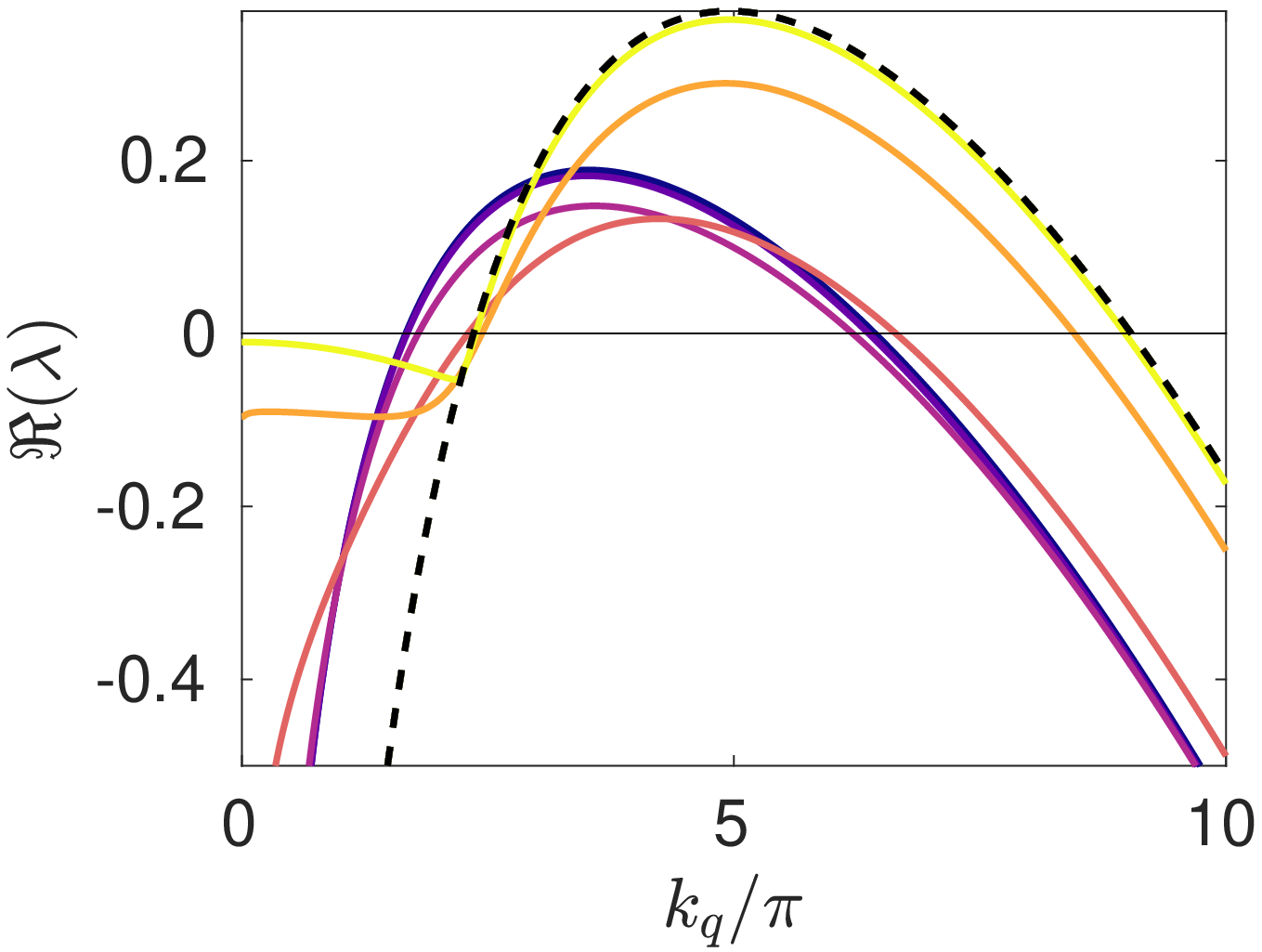}} \\
    \subfloat[Case III: $h=3\times 10^{-2}, \varepsilon=10^{-3}$]{\includegraphics[width=0.45\textwidth]{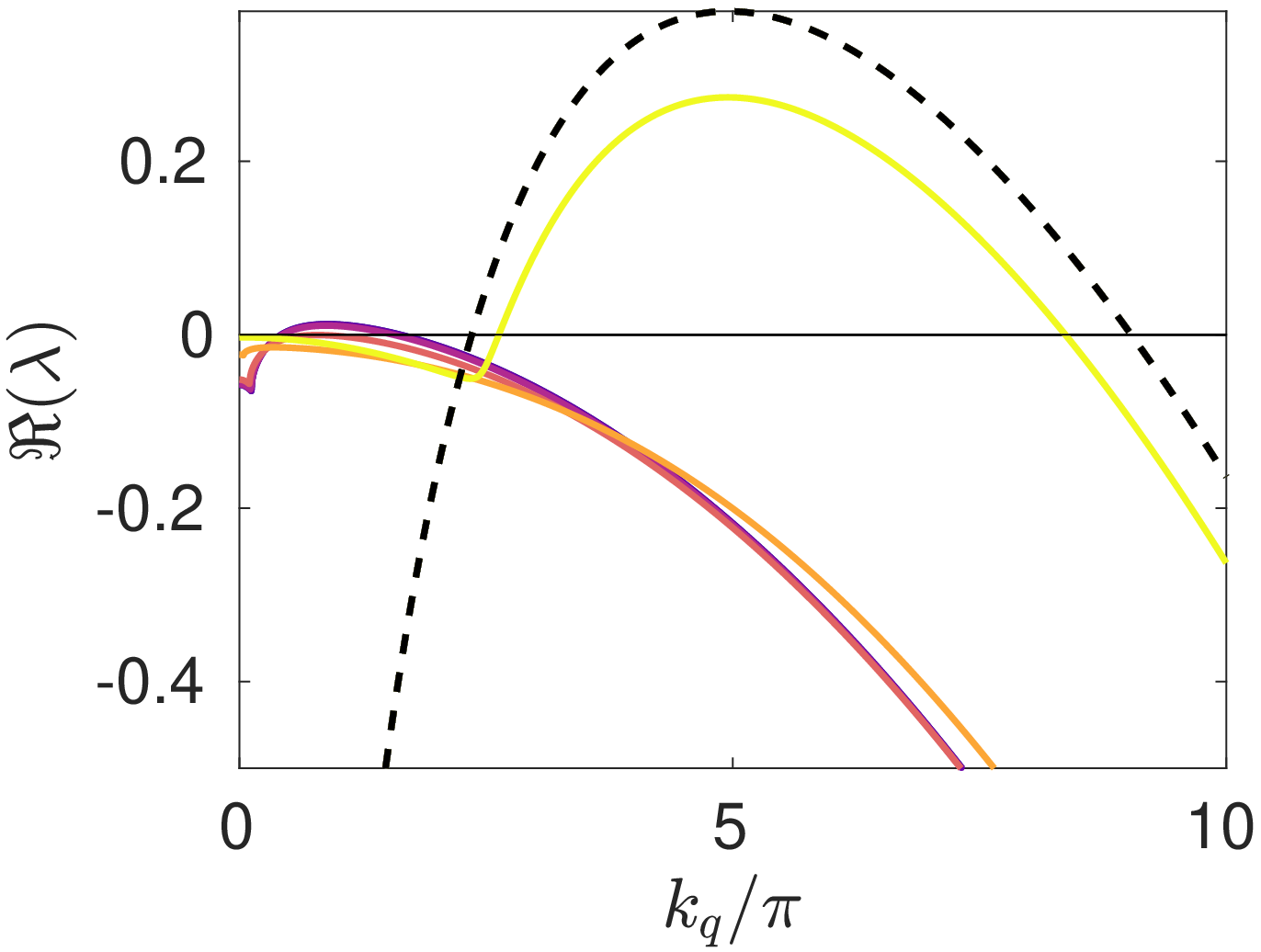}} &
    \subfloat[Case III: $h=3\times 10^{-2}, \varepsilon=10^{-3}$, $\bm{D_B}=\bm{I_2}$]{\includegraphics[width=0.45\textwidth]{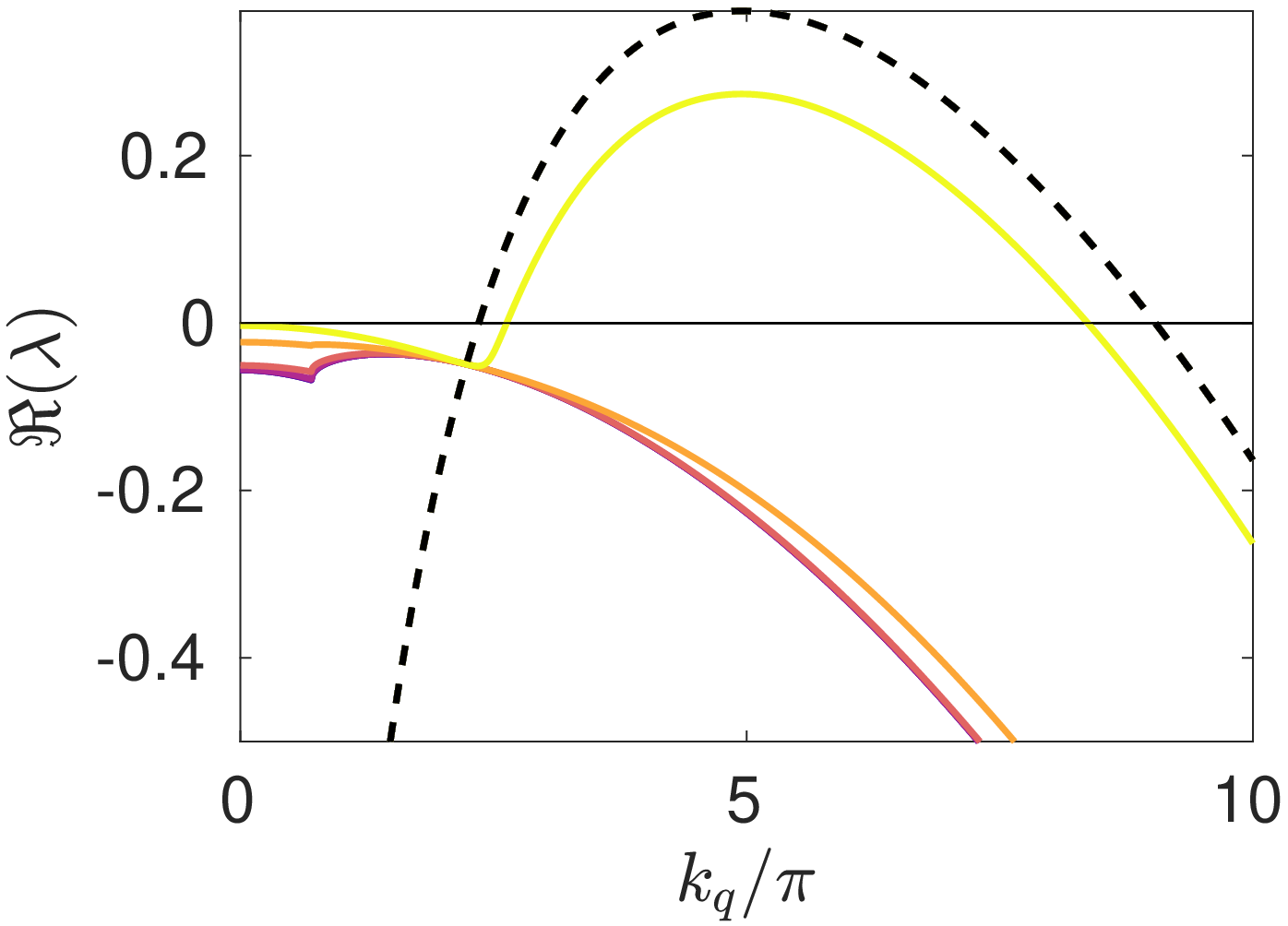}}
  \end{tabular}
  \includegraphics[width=0.9\textwidth]{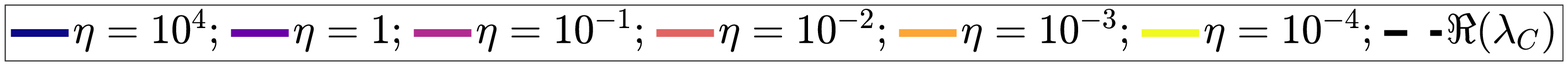}
  \caption{ Dispersion relations in the $x$  coordinate  computed via \eqref{detconfFull} for a continuous variable $k_q$, using the parameters $a=0.1$, $b=2$, with surface diffusion parameters $d_u=10^{-3}$, and $d_v=10^{-1}$. In (a) -- (c) we take $\bm{D_B}=\bm{D_S}$, though in (d) we set $\bm{D_B}=d_u\bm{I_2}$, corresponding to equal bulk diffusion between species. The solid lines correspond to $\max(\Re(\lambda))$ for different values of $\eta$ for the bulk-surface condition, whereas the dashed line corresponds to the single-domain classical case. For (c) we  anticipate there is an instability for relatively low $k_q/\pi$ and large $\eta$ due to  surface-bulk interaction instabilities, as exemplified in Section~\ref{ftsar} for $h  \sim \mbox{ord}(\varepsilon^{1/2})$ and $ \varepsilon_*^{2}/(3\varepsilon^{1/2} ) \ll 1$.} 
  \label{dispfig2}
\end{figure}


We first consider numerical constructions of dispersion relations for the small-asymptotic limits described in the previous section. Here we consider $\Re(\lambda)$ as a continuous function of the spectral parameter in the $x$ direction, $k_q$, as is commonly done \cite{Murray2003}. For  $\varepsilon, \varepsilon_* \ll 1$, we have that the isolated reaction-diffusion system can admit growth rates $\lambda$ comparable to a classical one dimensional reaction-diffusion system, which we will denote by $\lambda_C$ (which can be computed in the standard way \cite{Murray2003}). We can then consider the maximum value of $\Re(\lambda)$ (across all values of $\lambda$ found from condition \eqref{detconfFull}), and compare this to the isolated case.  We have confirmed these dispersion relations against full numerical simulations by simulating on a domain of lateral size $\tilde{L}$ such that a particular mode $k_q = q\pi/\tilde{L}$ is admissible, and observing a patterned solution. 

We plot these dispersion curves in Figure~\ref{dispfig2} for a variety of the geometric and coupling parameters. As anticipated, the coupling strength $\eta$ and geometric parameters $h, h_*$ and $\varepsilon, \varepsilon_*$ each influence the shape of these dispersion curves greatly. We now compare these curves to the predictions in Table~\ref{tablelims}. For Case I ($h \ll \varepsilon$), we see that  $\max(\Re(\lambda))$ is almost unchanged to the standard case up to small corrections not captured by the asymptotics. In Case II ($h \sim \varepsilon$), we observe approximate equivalence of the dispersion curve to the isolated case for small $\eta$, and an apparent change in the dispersion relation for increasing $\eta$. The Case III behaviour ($h \gg \varepsilon$) is consistent with the asymptotics of  Table~\ref{tablelims} whenever $\Re(\lambda)>0$ except  for  $\eta \sim \varepsilon \|{\bm J_S}\|_\infty$ and $\eta \gg \varepsilon\|{\bm J_S}\|_\infty$ at relatively small values of $k_q/\pi$.  Given these constraints, this  mismatch is anticipated   to be due to the   interaction between the surface kinetics and the bulk diffusion, as described in  Section~\ref{ftsar} given  the thin surface approximation
$\varepsilon_*^2/(3\epsilon^{1/2})\ll 1$ with  $k_q^2||\bm{D_S}||_\infty, |\lambda|\sim$ord$(\varepsilon^{1/2}||\bm{J_S}||_\infty)$. As a consistency test of this suggested mechanism,  in Figure~\ref{dispfig2}(d) we replace $\bm{D_B}$ by a scaled identity matrix so that differential diffusion in the bulk is no longer present, and we see that all of the dispersion curves, for smaller values of $k_q/\pi$ and $\eta$ sufficiently large, fall below the axis as expected. This is true for different scalar multiples of the identity, such as $\bm{D_B} = d_v\bm{I_2}$ where the dispersion curves were even more stable. We remark that considering other parameters demonstrates that this nontrivial bulk-surface interaction can lead to a non-monotonic behaviour of the dispersion relation with respect to $\eta$.
 
As in the classical case, we expect that for sufficiently large domains, any region where $\Re(\lambda)>0$ should admit a patterned state. We confirmed this using $\tilde{L}=100$ for each of the dispersion curves, finding that they admitted patterned solutions for long time simulations if and only if $\Re(\lambda)>0$ for some region in $k_q$-space. Similar to the classical case, the layered model is always observed to  stable at $k_q=0$ though with a local maximum at this point, in contrast to the behaviour of the classical Turing instability  dispersion relation.

To compare these dispersion relations against numerical simulations of the full nonlinear system, we compute a heterogeneity functional determining how far a solution is from a homogeneous state \cite{berding1987heterogeneity}. For simplicity, and because the surface layer is of 
primary interest in synthetic pattern formation within
bacterial colonies, we  only consider the heterogeneity of the activator in the surface.  We define the heterogeneity functional as 
\begin{equation}\label{hetero}
    F_h (u_S) = c\int_0^1 \int_h^{h+\varepsilon} \left(\frac{\partial u_S}{\partial x}\right)^2+\left(\frac{\partial u_S}{\partial y}\right)^2 \mathrm{d}y\mathrm{d}x,
\end{equation}
where $c >0$ is  simply  a positive definite (dimensional) scaling parameter. Note that $F_h(u_S) \geq 0$ and for $u_S \in C^1$, $F(u_S) = 0$ if and only if $u_S$ is spatially homogeneous. While we do not anticipate this metric to be quantitatively comparable to $\max(\Re(\lambda))$, we note that near the boundary of a Turing instability, the amplitude of patterns and their growth rates in time both scale with the distance from the bifurcation point, typically as a square root of the growth rate \cite{cross1993pattern}. Hence this functional should at least qualitatively scale with the growth of $\max(\Re(\lambda))$ near the onset of instability. The value $c$ is taken so that $F_h(u_S)=\max(\Re(\lambda))$ when $\eta=0$ for scaling purposes. We note that these plots are intended to demonstrate qualitative, rather than quantitative, behaviour near the onset of instability. In particular, we anticipate quantitative disagreement between $F_h(u_S)$ and $\max(\Re(\lambda))$ when $\eta$ is large, though the functional will still indicate whether or not $\max(\Re(\lambda))$ predicts pattern formation, as well as the scaling of pattern heterogeneity as a function of $\max(\Re(\lambda))$ near the onset of instability.

To use this heterogeneity functional, the full system \eqref{eqS}-\eqref{coupling} was solved until a final time of $t=10^5$ to ensure a good representation of the steady state pattern. The initial data were taken to be $u_0 = {u^*}(1+\xi_u(x,y))$ and $v_0 = {v^*}(1+\xi_v(x,y))$ with $\xi_u$ and $\xi_v$ random fields such that at each value of $(x,y)$, they are independently and identically distributed normal random variables with zero mean and variance $10^{-4}$. The equations were simulated using the COMSOL Multiphysics\textsuperscript{\tiny\textregistered} software \cite{COMSOL} with at least $2\times 10^4$ second-order triangular finite elements. A non-uniform mesh was constructed such that the surface region $\Omega_S$ was resolved with at least $10$ distinct triangular elements in any vertical cross-section. Convergence was checked in spatial and temporal discretisations, and a relative tolerance of $10^{-5}$ was given to the adaptive timestepping algorithm.

\begin{figure} \centering
  \begin{tabular}{cc}
    \subfloat[$h=1, \varepsilon=10^{-3}$]{\includegraphics[width=0.4\textwidth]{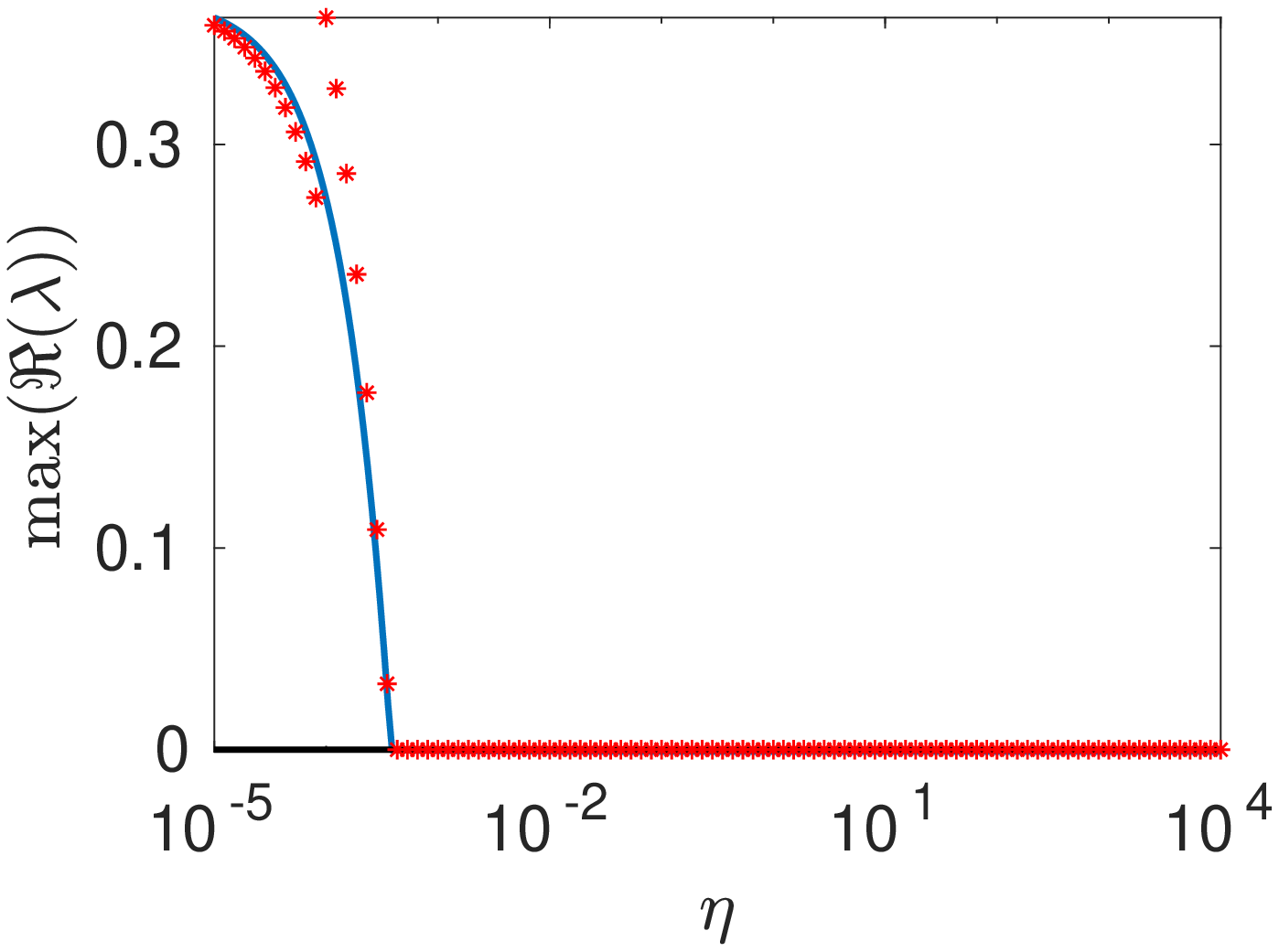}} &
    \subfloat[$h=1, \varepsilon=10^{-2}$]{\includegraphics[width=0.4\textwidth]{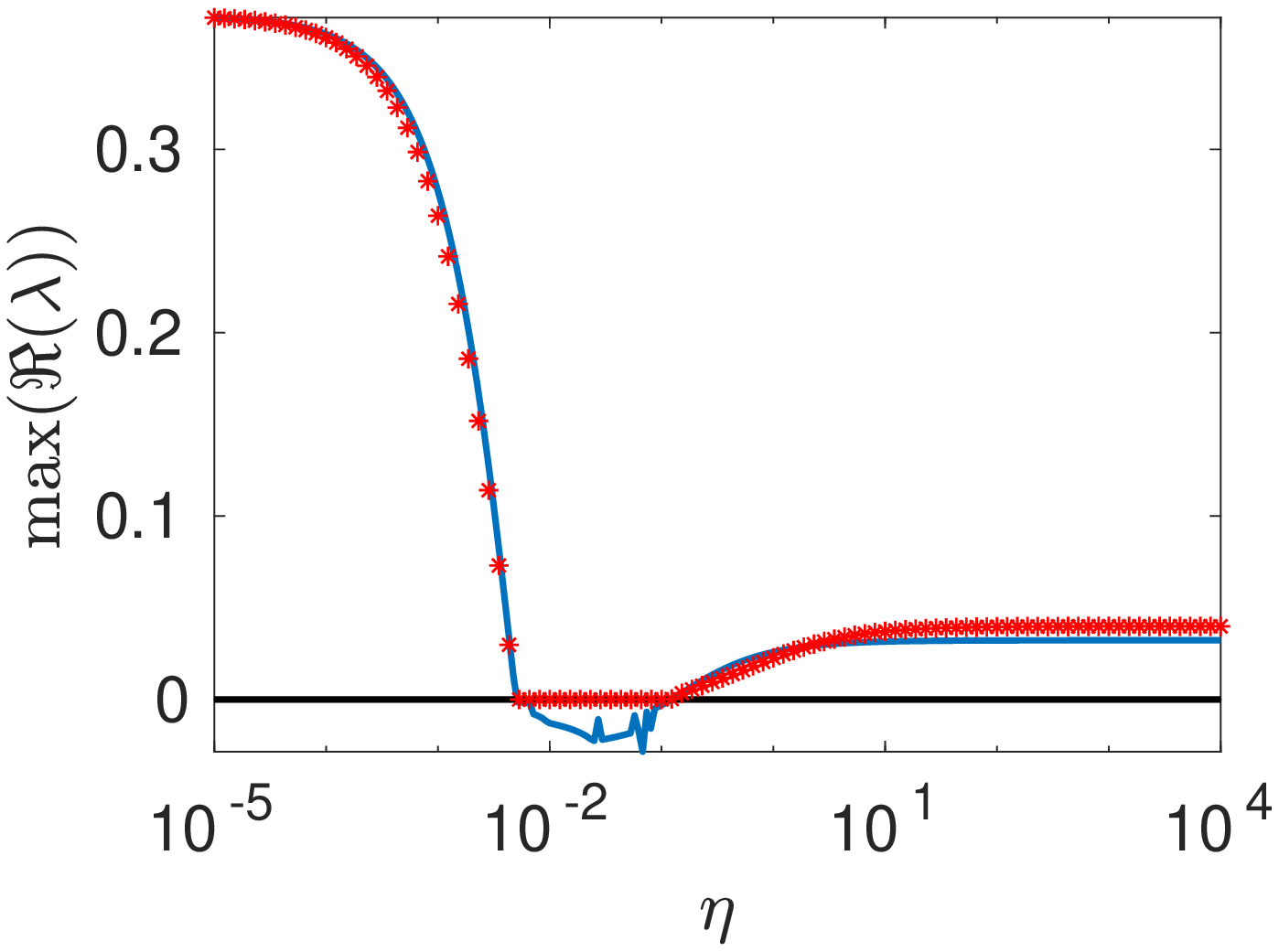}} \\
    \subfloat[$h=10^{-2}, \varepsilon=10^{-2}$]{\includegraphics[width=0.4\textwidth]{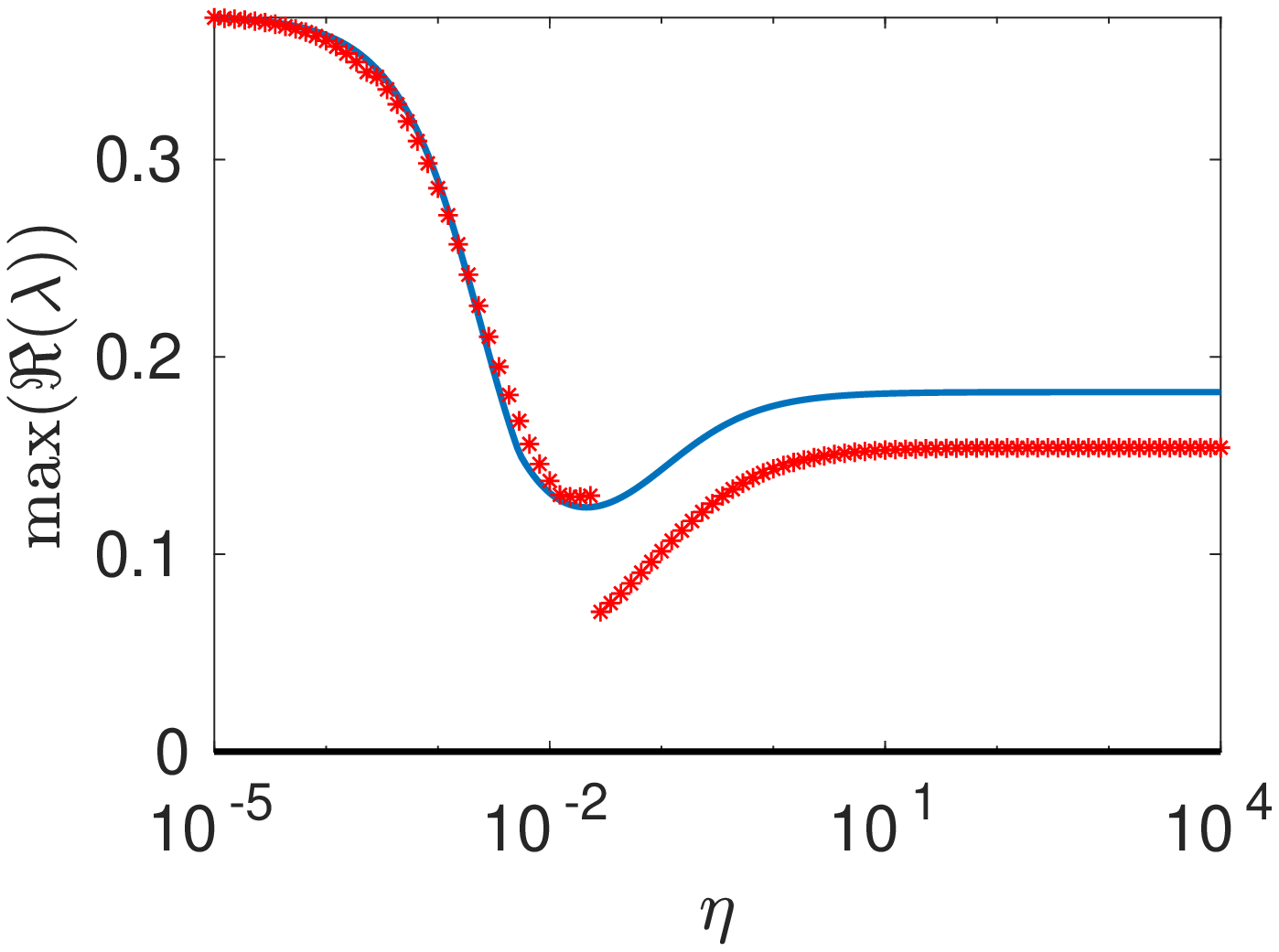}} &
    \subfloat[$h=10^{-1}, \varepsilon=10^{-2}$]{\includegraphics[width=0.4\textwidth]{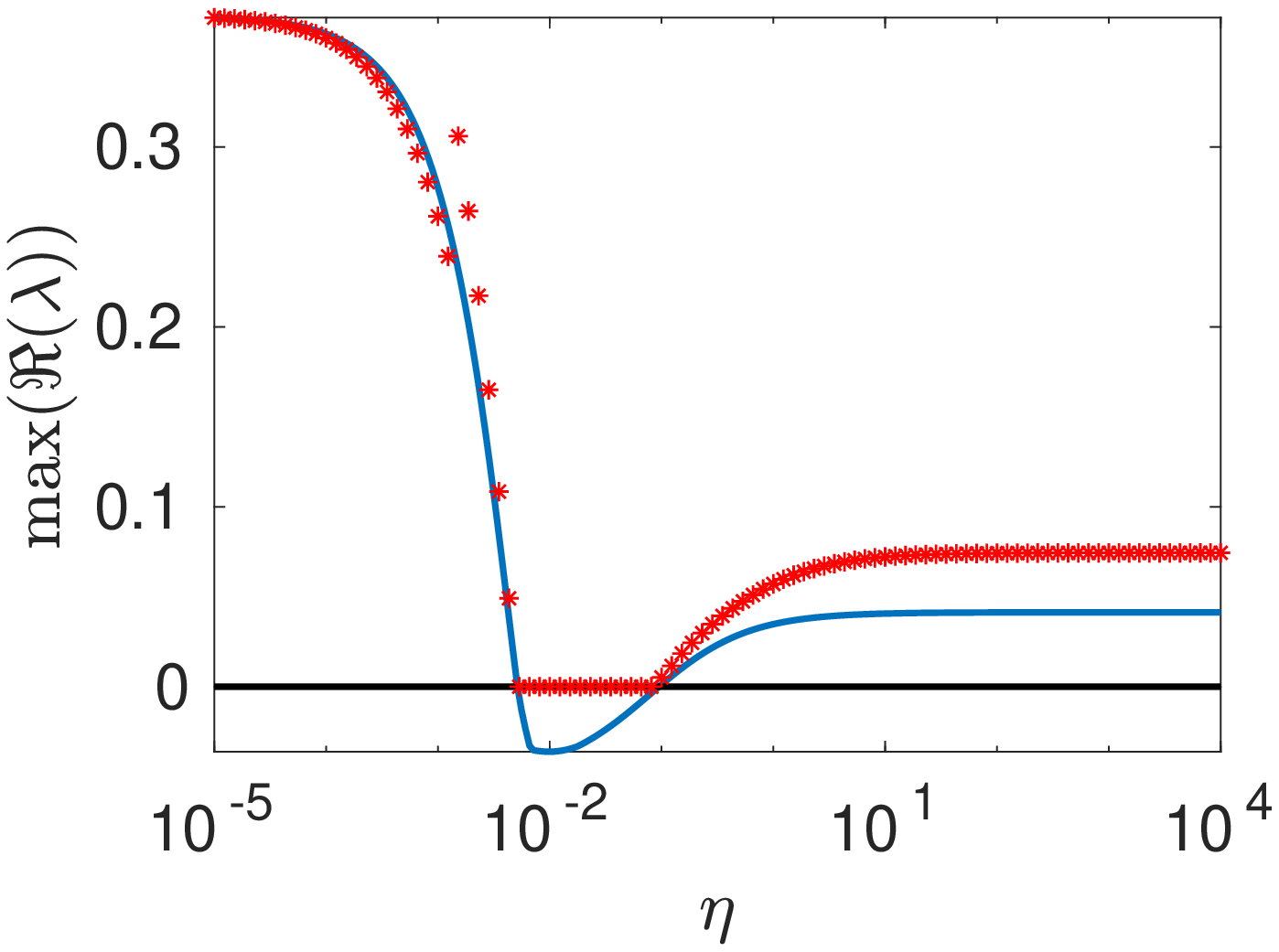}} \\
  \end{tabular}
  \caption{Non-trivial dependency of Turing instabilities on geometric parameters. Plots of $\max(\Re(\lambda))$ given by equation~\eqref{detconfFull} in blue computed across 250 values of $\eta$ for different parameter combinations, and plots of $F_h(u_S)$ given by \eqref{hetero} in red asterisks for 100 values of $\eta$. The other parameters were taken as $a=0.1$, $b=2$, $d_u=10^{-3}$, $d_v=10^{-1}$, $\tilde{L}=1$. The constant $c$ in $F_h$ was fixed per set of parameters/panel to match the maxima of $\max(\Re(\lambda))$ and $F_h$ across $\eta$ to qualitatively compare these metrics. The parameter sets corresponding to $h=10^{-1}, \varepsilon=10^{-1}$ and $h=1, \varepsilon=10^{-1}$ gave qualitatively the same results as in panel (c) with $\max(\Re(\lambda))>0$ for all $\eta$. }
  \label{stabfig}
\end{figure}

\begin{figure} \centering
  \includegraphics[width=0.8\textwidth]{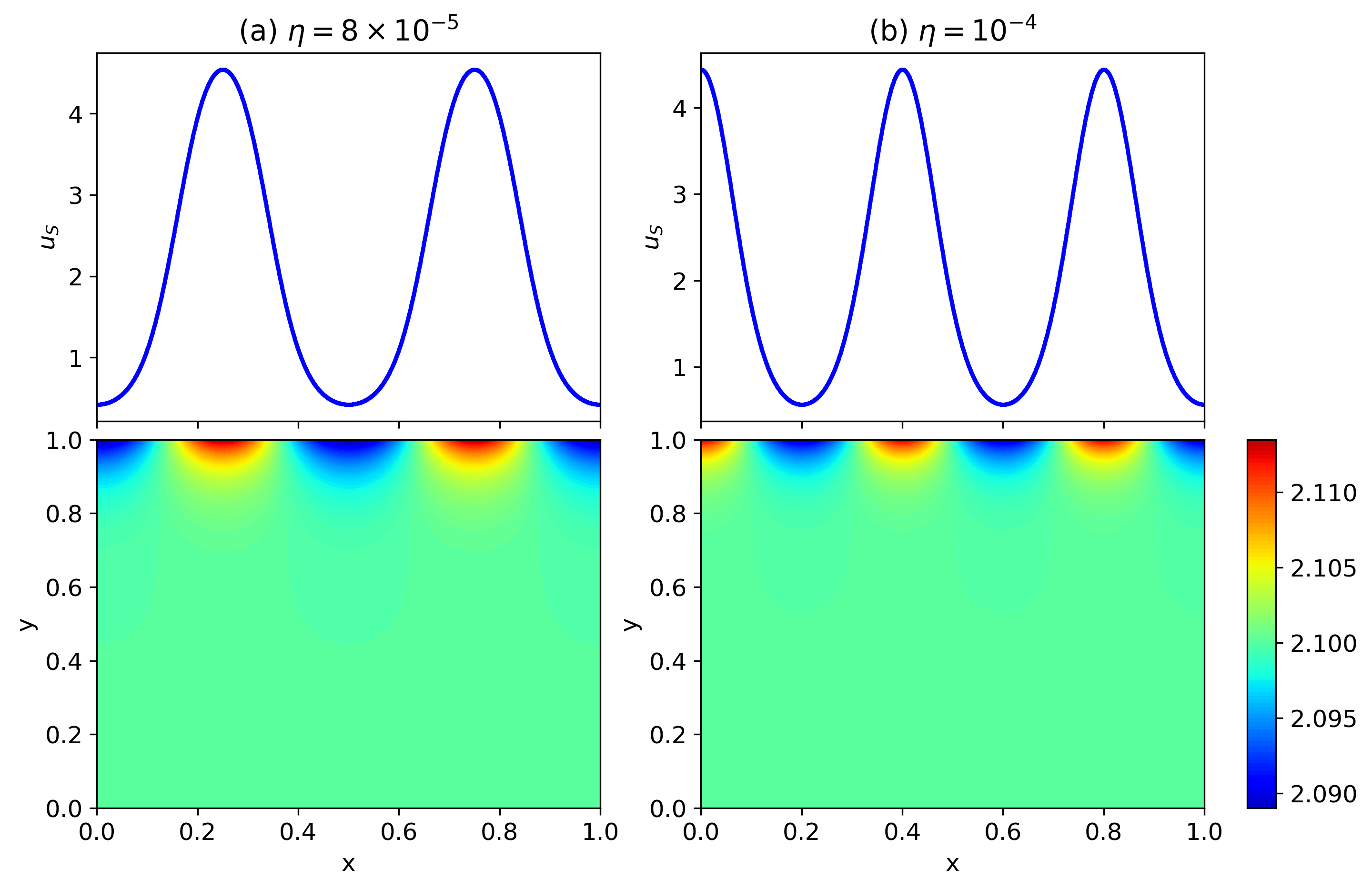}
  \caption{One-dimensional plots of $u_S$  corresponding to parameters in Figure~\ref{stabfig}(a) for two values of $\eta$ in the top two panels, and plots of the corresponding $u_B$ below (with $\tilde{L}=1$ in all cases). The surface concentration $u_S$ is effectively homogeneous in the $y$ direction, and so is essentially a one-dimensional pattern, shown above. Note that the bulk concentrations are almost homogeneous, whereas the surface concentrations are not (compare the scales of $u_S$ and $u_B$). }
  \label{h1eps1e-3}
\end{figure}

In Figure~\ref{stabfig} we give examples of this heterogeneity functional across the ranges of the geometric parameters $\varepsilon$, $h$, and $\eta$, alongside predictions from the instability condition \eqref{detconfFull}. As anticipated by the asymptotics, for very small $\varepsilon$ (Figure~\ref{stabfig}(a)), we see the system fails to support spatial patterns for $\eta \geq 3.9 \times 10^{-4}$. Additionally, we see a jump in the value of the heterogeneity between $\eta = 8\times 10^{-5}$ and $\eta = 10^{-4}$. We plot values of $u_B$ in Figure~\ref{h1eps1e-3} across this jump to demonstrate that this discontinuity in the value of the spatial heterogeneity $F_h(u_S)$ for these parameters is due to different nonlinear modes emerging as parameters are varied, and so it is sensible that it is not captured in the linear analysis. Other discontinuities in the plots of the heterogeneity functional in Figure~\ref{stabfig} are similarly due to different patterned states being selected, and we do not further explore pattern multistability or dependence on initial data here. 

\begin{figure} \centering
  
  
    \includegraphics[width=\textwidth]{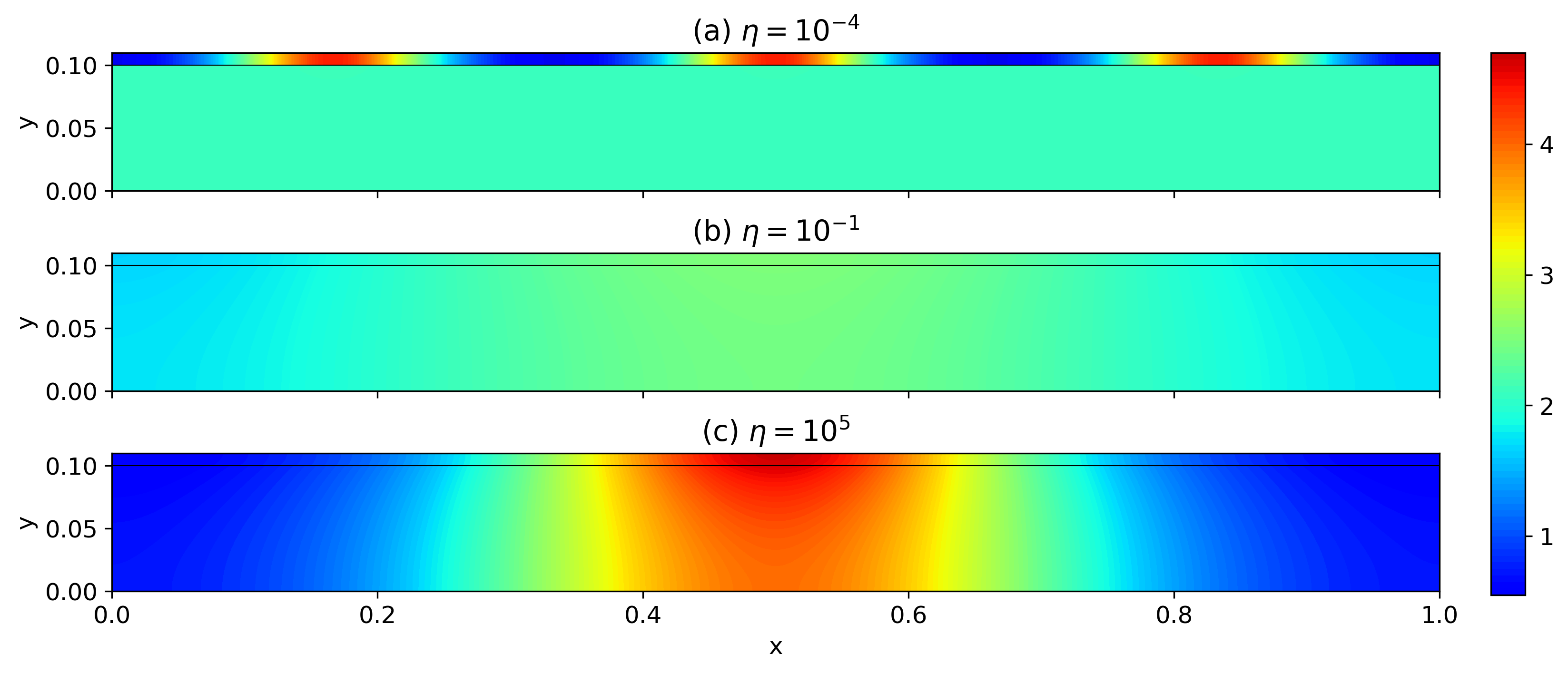}
    \caption{Plots of $u_S$ and $u_B$ corresponding to parameters in Figure~\ref{stabfig}(d) for three values of $\eta$, and $\tilde{L}=1$. Here, $\varepsilon=10^{-2}$ and $h=10^{-1}$.}
    \label{h0p1eps1e-2}
\end{figure}

In Figures \ref{stabfig}(b) and (d) we see a region of intermediate values of $\eta$ for which no patterning occurs, and more broadly across all of Figure~\ref{stabfig} we see that a minimal value of $\max(\Re(\lambda))$ occurs approximately for $\eta$ within the range $ (10^{-3}, 1)$. We show examples of the mode selection process from Figure~\ref{stabfig}(d) in Figure~\ref{h0p1eps1e-2}.  For small $\eta=10^{-4}$, we see stable  multiple-spike solutions that are  essentially confined to the surface.   As $\eta$ increases further to $10^{-1}$, a single-spike solution is observed, at a smaller amplitude as the dispersion relation has just crossed the instability threshold given in Figure~\ref{stabfig}(d). Further increases to large $\eta$ lead to stable spike solutions that remain essentially vertically homogeneous in the surface, but have small transverse variations in the bulk due to the change in reaction kinetics across the interface, as illustrated for $\eta=10^5$.  Further increasing $\eta$ sharpens these spike solutions across the domain, but does not impact the number of modes. Besides the discontinuities in the heterogeneity due to nonlinear mode selection, there is often a good match between the linear analysis (e.g.~value of $\max(\Re(\lambda)$) and the heterogeneity, which can be expected near to the Turing bifurcation points in simpler settings due to the existence of normal forms of the pattern amplitude \cite{cross1993pattern}. 

In all of Figure~\ref{stabfig} we observe that $\max(\Re(\lambda))$ appears to asymptotically approach a fixed value for either $\eta \to 0$ (which corresponds to the static Turing conditions) or $\eta \to \infty$, with the latter always being smaller than the former, though this may just be a feature of the parameters explored here. However, in \ref{stabfig}(c) (and the other cases noted in the caption), we observe that an instability occurs for all values of $\eta$, which is confirmed by numerical simulations of the full system. 

\begin{figure} \centering
    \includegraphics[width=\textwidth]{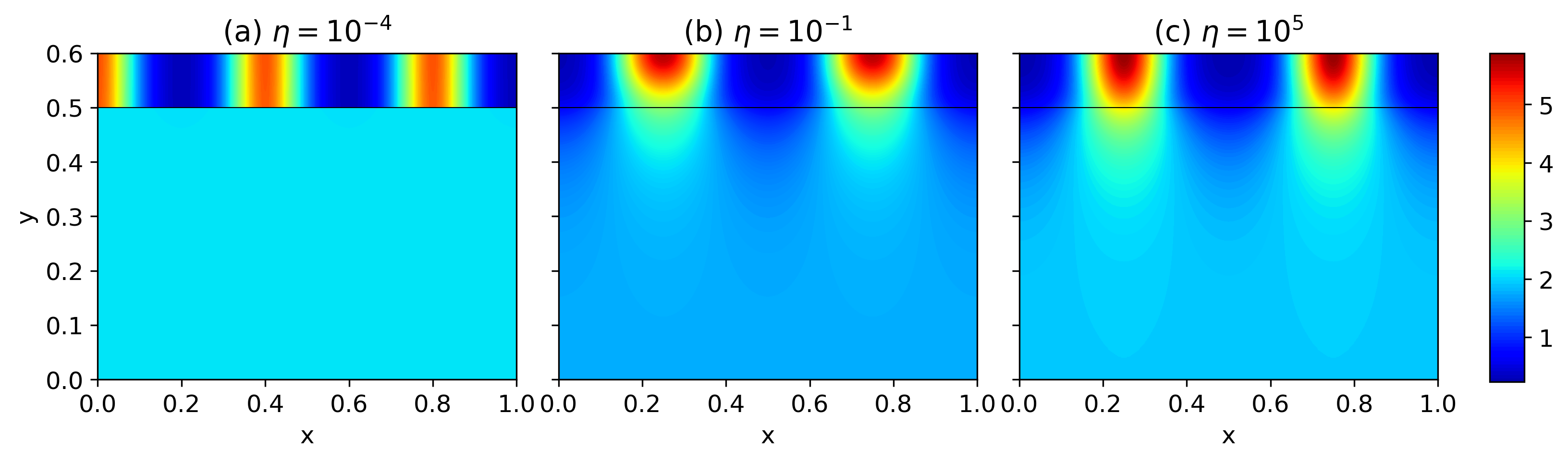}
    \caption{Plots of $u_S$ and $u_B$ corresponding to parameters in Figure~\ref{stabfig} with $h=0.5$ and $\varepsilon=10^{-1}$ for three values of $\eta$, and $\tilde{L}=1$. }
    \label{h1eps0p1}
\end{figure}

\begin{figure} \centering
    
    
    \includegraphics[width=\textwidth]{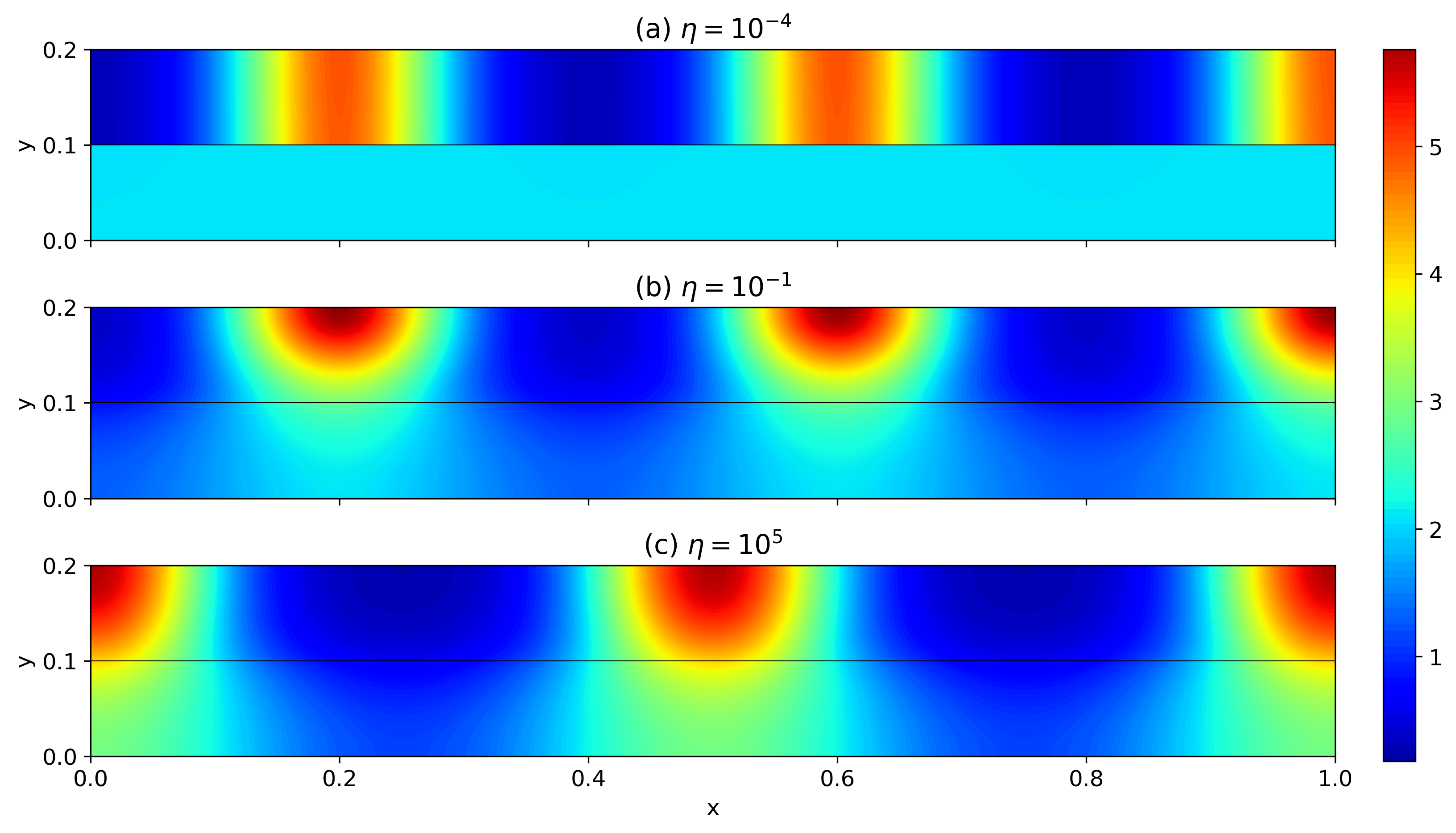}
    \caption{Plots of $u_S$ and $u_B$ corresponding to parameters in Figure~\ref{stabfig} except that $h=10^{-1}$ and $\varepsilon=10^{-1}$ for three values of $\eta$, and $\tilde{L}=1$.}
    \label{h0p1eps0p1}
\end{figure}

As a further example which helps visualise the impact of varying the geometric parameters and coupling constant $\eta$, we observe patterns primarily confined to the surface but with some interaction with the bulk in Figure~\ref{h1eps0p1}. Again some mode selection effects are present (two vs three spot solutions for small and largeer values of $\eta$ respectively), though due to generic aspects of multistability in two spatial dimensional systems \cite{dewel1995pattern}, we suspect these depend somewhat on initial data, rather than just parameter values. Finally in Figure~\ref{h0p1eps0p1} we give an example where no change in the number of unstable modes was apparent for variation in $\eta$, though the structure of the solution does change.    

Within Turing-unstable regimes, the surface largely drives the structure of the modes and hence the patterns can be thought of as quasi-one-dimensional (Figures \ref{h1eps0p1}-\ref{h0p1eps0p1}). The permeability $\eta$ does control how much structure there is, both in the bulk in general and in the surface modes' variation in the $y$ direction, though in all cases the largest spatial variation is along the lateral coordinate $x$. For the largest permeability we explored ($\eta=10^5$), we see that the sizes of the surface and bulk can have a significant impact on the relative shape of the solutions in the bulk region (cf Figures \ref{h0p1eps1e-2}(c), \ref{h1eps0p1}(c), and \ref{h0p1eps0p1}(c)). In particular we see that the deepest part of the bulk ($y=0$) in Figure \ref{h0p1eps1e-2}(c) and \ref{h0p1eps0p1}(c) maintain a fairly distinct periodic pattern between high and low activator concentrations, whereas the larger bulk in Figure \ref{h1eps0p1}(c) is substantially more homogeneous at $y=0$. We also note that for intermediate values of $\eta$, Figures \ref{h1eps0p1}(b) and \ref{h0p1eps0p1}(b) have the largest visual gradients in the activator in the surface layer, consistent with the intermediate-$\eta$ values having significant impacts on the predicted values of $\max(\Re{\lambda})$ in Figure \ref{dispfig2}. This further demonstrates nontrivial impacts of the bulk geometry on the structure of emergent patterns, and such leeching into the bulk may be useful to help quantify its impact in synthetic systems.

\section{Discussion}\label{Discussion}

Motivated by recent interest in a range of biological contexts, we have developed and analysed a general class of reaction-diffusion models of pattern formation in stratified media, though with an absence of reactions in the bulk and a linear coupling between the layers. We have derived a criterion for pattern-forming instability in such media, given by equation~\eqref{detcond}. In Appendix \ref{App} we showed that the absence of differential transport within each layer entails no patterning for these systems, in direct analogy to the classical Turing instability. We have also demonstrated a range of interesting behaviours via asymptotic reductions in thin domains, and numerical simulations.
In particular, this setting of a linearly coupled system with no reactions in the bulk with a thin surface layer is also of significant biological interest, as several groups are using bacterial colonies on inert substrates as a medium for engineered pattern formation via synthetic biology \cite{grant2016orthogonal,boehm2018,karig2018stochastic}. However,  as far as we are aware, there is little theoretical understanding of how the inert substrate impacts the surface reaction-diffusion systems in these kinds of geometries. Additionally, to accurately model the real complexity of these experimental systems we would need to account for intracellular (i.e.~non-diffusible) proteins which play a role in the reactions, as our reaction-diffusion framework only captures the dynamics of diffusible signalling molecules.

Nevertheless, even in the simplified setting of an inert bulk and a thin surface, the  computed instability criteria are much richer than in the classical case. For instance,  the nine distinguished limits for $|\lambda|, k_q^2\|\bm{D_S}\|_\infty \sim \mbox{ord}(\| \bm{J_S} \|_\infty)$  given in Table~\ref{tablelims} demonstrate a variety of behaviours not predicted by analysing the surface reaction-diffusion system alone, as is typical in applications. In addition, these distinguished limits, though emergent from a complex multi-parameter system, depend on only three non-dimensional parameter groupings, $\varepsilon_*,h_*$ and $\eta/(  \varepsilon \| \bm{J_S} \|_\infty )$. The first two of these respectively are the surface and bulk depth relative to the lateral lengthscale, i.e. the basic geometry.  The final grouping is $\eta/(  \varepsilon \| \bm{J_S} \|_\infty )=\tau \hat{\eta}/(H_\varepsilon \| \bm{J_S} \|_\infty )$. 
Noting that  $\tau$ is chosen such that $\| \bm{J_S} \|_\infty\sim\mbox{ord}(1)$, one can deduce more generally that $\tau/\| \bm{J_S} \|_\infty$ is the dimensional timescale of surface reaction. 
Hence the final parameter grouping is the ratio of the interface  permeability to the surface velocity scale, $\| \bm{J_S} \|_\infty H_\varepsilon/\tau$, with the latter in turn given by the ratio of the surface depth and reaction timescale. 

We further note that our instability condition \eqref{detconfFull}, recovers the usual features of Turing instabilities, such as requiring differential diffusion for their onset, and reducing to the polynomial dispersion relation when the bulk becomes uncoupled from the surface. The explicit coupling between bulk diffusion and surface reactions given by \eqref{smallepslambdakq2} when  $|\lambda|, k_q^2\|\bm{D_S}\|_\infty  \sim \mbox{ord}(\varepsilon^{1/2}\| \bm{J_S} \|_\infty)$  suggests   additional distinguished limits from those in Table~\ref{tablelims};  the associated instabilities possess  slower  growth rates, but nonetheless highlight substantial and non-trivial impacts of the bulk on the system's ability to pattern.  We anticipate that there are other examples of nontrivial surface-bulk coupling driven instabilities, as suggested in the discussion of the Averaged and Quadratic cases in Table~\ref{tablelims}, but leave investigation of these to further work.

Broadly, our asymptotic and numerical results on thin surfaces suggest that the presence of the inert bulk generally decreases the ability of the surface system to undergo a Turing instability compared to an isolated system. The exceptional cases, such as the homogenised limit \eqref{smallepshleta} and the explicit coupling in equation~\eqref{smallepslambdakq2}, can in principle lead to larger Turing spaces, though we have shown in some realistic cases such as equal bulk diffusions ($\bm{D_B}=\bm{I_n}$) that these do not enlarge the Turing space. Note that in systems where diffusion varies significantly between domains (e.g. non-diffusible proteins in the surface) the parameter space that admits pattern formation can increase with increasing bulk size (see, for instance \cite{halatek2018box, brauns2020bulk}). Exploring such interplays will be the focus of future work.

Our results suggest that experiments should aim to design large and robust parameter regimes using classical criteria for pattern-formation (e.g.~using design approaches such as in \cite{dalchau2012towards}), as diffusion into the bulk region will likely decrease the size of such Turing spaces. We have shown that even in cases where the broad influence of the bulk is to decrease the ability of the system to pattern, such a decrease will be non-monotonic in the geometric and transport parameters of the bulk region in general, as illustrated with the non-dimensional bulk depth, $h$  and permeability, $\eta$. Many of the parameters may not be controllable, though one can often choose an agar height $h$ above a certain minimal threshold. The results in Table \ref{tablelims} broadly suggest that the agar layer should be made as thin as possible to limit the impact on a system's ability to pattern. There may also be opportunities to decrease the permeability into the bulk, $\eta$, by using thicker filter paper or modifying the pore size or density, which would also reduce the negative impact of the bulk on pattern formation, though due to metabolic constraints (as the agar is primarily a nutrient) this too may be somewhat limited.  We do note that there are important experimental controls in the genetic circuits encoded in the nonlinear reaction kinetics, which we have only caricatured in this study by considering the two-species case with only diffusible morphogens. Finally, we have shown instances of instability such that the bulk domain is a necessary component to drive an otherwise stable surface system to a patterned state (e.g.~equation~\eqref{smallepslambdakq2} and the following discussion), though we leave systematic analysis of such instabilities for future work. This route to instability does not contradict the preceding suggestions about reducing $\eta$ and $h$, as it is likely inadmissible for bacterial pattern formation on agar. Such an experimental setting entails that it is reasonable to assume $\bm{D_B} \propto \bm{I}$, and hence by \eqref{smallepslambdakq2} we see that bulk diffusion will not drive an instability in this case.

We remark that the mode selection phenomena we have illustrated (e.g.~in Figure \ref{h0p1eps1e-2}) can be understood in the context of finite-size effects, which are well-studied in the classical case \cite{Murray2003}. Namely given a dispersion relation for $\Re(\lambda_C(k_q))$,  where $\lambda_C$ is the growth rate of a classical Turing mode, one can tune the geometry to select different spatial eigenvalues $k_q$ to give rise to non-monotonic effects as, for instance, the domain size is increased. However, here the effects are more subtle as we cannot explicitly compute the relationship between $\lambda$ and eigenvalues of the full spatial operator, and so can only implicitly observe these effects. Nevertheless, these mode selection effects appear to be more prevalent compared to classical cases as they require very small domains and other fine-tuning \cite{Murray2003}. Additionally, in our setting mode selection effects appear to be more prevalent across a wide range of geometric parameters, whereas the classical cases have been studied almost entirely in terms of a scalar length, and are generally restricted in parameter regimes where they occur. In particular, we conjecture that the non-monotonic dependence of $\max(\Re(\lambda))$ on $\eta$ seen in Figure \ref{stabfig} is due to these effects, as we see different modes being excited on either side of this region in Figure \ref{h0p1eps1e-2}.

There are numerous extensions of these results that are worth pursuing. In the example setting of bacterial colony formation on an agar substrate, one might need to augment the bulk evolution with a degradation reaction. We remark that such a simple addition leads to substantial complexity as, if the surface equilibrium is nonzero, then there does not exist a homogeneous equilibrium across the whole coupled system (a degradation reaction in the bulk by itself will always lead to a homogeneous zero equilibrium concentration). We anticipate that the mathematical structure in this case will be even more intricate. 
A simpler addition, also of relevance to bacterial patterning on agar, would be the inclusion of non-diffusible reactants in the surface region. This approach would also pave the way to account for all gene regulatory dynamics in a quantitative model based on mass action kinetics.  In such a case, we can apply techniques to incorporate the impact of such reactants on the surface reaction kinetics directly (in the linearised system) \cite{klika2012influence}. 
Along similar lines, more complicated transport functions $\bm{g}$ across the membrane can be studied, again leading to new possibilities of differential transport, which can easily be added to the analysis implemented here. We have also assumed that the same number of species diffuse throughout both domains, but in principle one can generalize this by introducing different coupling functions $\bm{g}$ for the surface and bulk boundary conditions, presently given in equation \eqref{coupling}. Such an analysis is broadly similar though there are several key details to account for, so we leave this for further work.

There are many biological examples of physical layered media with reactions in multiple different spatial domains, such as in the epithelial-mesenchymal coupling during the development of the skin in mammals \cite{vilaca2019numerical}. For example, in the study of hair follicle morphogenesis, a substantial amount of biochemical research has implicated Turing-type instabilities in the formation of follicle primordia \cite{mou2006generation}. More recently, it has been suggested that a simple activator-inhibitor system is insufficient to capture the dynamical complexity in hair follicle patterning, and so suggestions have been made that such patterns arise due to many coupled processes, which will undoubtedly occur across the different domains of the epithelium and the developing mesenchyme \cite{glover2017hierarchical}. Similar remarks can be made about many kinds of skin and other organ patterning events across a range of species, suggesting that general methodologies for stratified reaction-diffusion systems would be useful to elucidate underlying physico-chemical mechanisms. A related layered system is the synthetic pattern formation studied in a monolayer of HEK293 cells grown beneath a culture medium \cite{sekine2018synthetic}, where presumably bulk diffusion plays a significant role in transporting signalling molecules. 

While we have explored exemplar reaction-diffusion systems in such coupled domains, there are more general transport mechanisms that could be studied. Both chemotaxis and a range of mechanical taxis, as well as mechanical forces, could be included in such a model. We note that a numerical study \cite{vilaca2019numerical} has made some progress towards such a model. The linear stability analysis for such problems is involved, but the approach presented here generalises to these settings. Of course, in the absence of a homogeneous steady state, one must develop new methods for the analysis of pattern-forming instabilities. This has been done recently for heterogeneous steady states \cite{krause_WKB}, but extending such an analysis to these coupled geometries is nontrivial. Mathematically, the limit of $\eta \to \infty$ can be thought of as a step function heterogeneity, as explored in \cite{stephetero}, so that the systems studied here are also in some sense a generalisation of piecewise-constant reaction-diffusion problems, providing another perspective on heterogeneous reaction-diffusion systems.

Another related generalization would be to study discrete or hybrid discrete-continuum formulations of these kinds of layered media, such as the recent hybrid Turing-type model proposed in \cite{macfarlane2020hybrid}. Turing's original paper contained a study of discrete cells \cite{turing1952chemical}, which was later extended in \cite{othmer1971instability} and more recently in \cite{nakao2010turing} to reaction-diffusion systems on discrete networks. Such a formulation has been extended to consider multiplex networks, themselves a model of discrete layered media \cite{gomez2013diffusion}, within which Turing pattern formation has also been studied \cite{asllani2014turing, kouvaris2015pattern}. Such systems deserve exploration on their own, in addition to relating them to spatially continuous analogues of Turing systems in stratified media.

Finally we mention that one could generalise from our setting of two planar domains to many more coupled domains, or to more complicated geometric settings, including those relevant for more realistic models of development, such as in the blastula stage or later stages of epithelial-mesenchymal development on complicated morphologies. While our approach may be generalisable to very different geometric settings, the dispersion relation we have found in this simple case is already somewhat difficult to analyse, and full numerical simulations may be more expedient. Nevertheless, analytically tractable results for this family of problems are valuable in understanding the role of coupled domain structures in pattern formation, as such scenarios are ubiquitous in biological settings.

\begin{acknowledgements}
A.L.K. and E.A.G. are grateful for support from BBSRC grant BB/N006097/1; V.K. is grateful for support from the European Regional Development Fund-Project ‘Center for Advanced Applied Science’ (no. CZ.02.1.01/0.0/0.0/16\_019/0000778) and the Mathematical Institute at the University of Oxford. In compliance with BBSRC's open access initiative, the data in this paper is available from http://dx.doi.org/xx.xxxx/xxxxxxxxxxxxxxxxxx.
\end{acknowledgements}

\appendix   
\section{Further analysis of possible patterning instabilities: surface isolation and equal diffusion coefficients} \label{App}
Here, we compute instability conditions from equation~\eqref{detconfFull} for further distinguished limits, in particular   (i) the limiting of decoupling the interaction of the surface and bulk regions, that is for sufficiently small $\eta \ll 1$ and   (ii) the limit of equal diffusion coefficients in each region for classical Turing patterning systems. In particular, (i) provides a useful consistency check of the modelling framework, while  (ii) confirms that  in the absence of differential transport  within at least one layer or between them,   patterning cannot occur for classical Turing systems, in direct analogy to the behaviour of the single layer classical Turing instability.

\subsection{Patterning in the limit of an isolated surface system, via sufficiently small non-dimensionalised permeability} \label{AppA}
We consider the patterning conditions for the isolated surface system in the case that $\eta \to 0$. 
Here,   we denote the eigenvalues of $\bm{M_{B}}$ as $\mu_{Bp}$, the eigenvalues of $\bm{M_{S}}$ as $\mu_{Sp}$, and the eigenvalues of $\bm{J_{S}}$ as $\nu_p$, where $p = 1,2,\dots,n$ in all cases, where $n$ is again the number of reactants. By continuity of the determinant, equation~\eqref{detconfFull} reduces for sufficiently small $\eta$ to the condition 
\begin{equation}\label{eta0}
\det(\bm{M_{B}}\tanh(h\bm{M_{B}}))\det(\bm{M_{S}}\sinh(\varepsilon \bm{M_{S}}))=0,
\end{equation}
so that the bulk and surface components decouple. As these hyperbolic trigonometric functions are at worst meromorphic,  
 the  zero of the determinants occur for    eigenvalues of the matrices $\bm{M_{B}},~\bm{M_{S}}$ that, respectively, are  roots of   $z\tanh(hz)=0,~z\sinh(\varepsilon z)=0$ for $z\in \mathbb{C}.$  Hence, equation~\eqref{eta0} is satisfied whenever $\mu_{Bp}=j{\rm i}\pi/h$, or $\mu_{Sp}=j{\rm i}\pi/\varepsilon$, with $j\in\{0,1,2,\ldots\}$,  natural. 

For the bulk component, this implies that $\det(\bm{M_{B}}^2+(j\pi/h)^2\bm{I_n})=0$ which, as this is a diagonal matrix, implies that the allowed growth rates are given by
\begin{equation}\label{MbAlone}
\lambda_p = -D_{Bp}(k_q^2+(j\pi/h)^2) = -D_{Bp}\pi^2(q^2+(j/h)^2) \leq  0 ,\, \text{for } q, j\in\{0,1,\ldots\},\,  p\in\{1,2 \ldots, n\},
\end{equation}
with $D_{Bp}$ denoting the $p$th component of the diffusion matrix $\bm{D_{B}}$, and recalling that  $k_q=q\pi/\tilde{L}$.
Hence these solutions do not drive an instability, noting that although these eigenvalues formally break the assumption we made that $\Re(\lambda) > 0$ used to deduce equation~\eqref{detconfFull}, we have equation~\eqref{eta0} is precisely \eqref{detcond} in the limit  $\eta\rightarrow0$, and, hence, we have not used this assumption to determine these eigenvalues.

Similarly, by the preceding discussion of the eigenvalues $\mu_{Sp}$ of $\bm{M_{S}}$, we have a condition for the existence of nontrivial eigenvectors for these eigenvalues given by 
\begin{equation}\label{surfaceTuring}
    \det(\bm{M_{S}}^2-\mu_{Sp}^2\bm{I_n}) = \det((q^2+j/\varepsilon)\pi^2\bm{I_n}+\bm{D_S}^{-1}(\lambda\bm{I_n}-\bm{J_{S}}))=0,
\end{equation}
which we recognise as exactly the condition one would find to compute $\lambda$ in a traditional $n$ species reaction-diffusion system posed on a rectangle with Neumann boundary data \cite{klika2012influence}. In practice finding all such solutions for a given eigenpair $q,j$ is a straightforward numerical problem. While these computations are more easily implemented by noting that the $\eta=0$ limit does separate into two uncoupled regions that can be analysed via standard methods, these results  serve as a useful  consistency check of equation~\eqref{detconfFull}.

\subsection{Identical Diffusion Coefficients Within Regions}\label{AppA2}
 Below we assume that the surface Jacobian, $\bm{J_S}$, can be 
 diagonalised, noting that diagonalisable matrices are a dense subset of complex-valued matrices  and thus the results derived below therefore hold in general due to continuity. In particular, we now show that if there is identical diffusion for  different species within each region, then the spatially homogeneous steady state is linearly stable to the perturbations given by equation~\eqref{expansions}  for kinetics in the surface layer that allow the classical Turing instability.

To proceed, we let $\bm{D_S} = c_S\bm{I_n}$ and $\bm{D_B} = c_B\bm{I_n}$, so that $\bm{M_{B}}^2 = (k_q^2+\lambda/c_B)\bm{I_n}$. As $\bm{J_{S}}$ can be diagonalised, we can also diagonalise $\bm{M_{S}}^2 = (k_q^2+\lambda/c_S)\bm{I_n}-(1/c_S)\bm{J_{S}}$. Making this additional assumption, we rewrite these matrices using the $n+1$ scalar values $m_B^2 = k_q^2+\lambda/c_B$ and $m_p^2 = k_q^2+\lambda/c_S-\nu_p/c_S$, where $\nu_p$ are the eigenvalues of $\bm{J_s}$ and $p=1,2,\dots,n$. We note that these scalars will never be zero as $k_q^2$ is real and non-negative, we require $\Re(\lambda) > 0$ and,  restricting ourselves to kinetics that exhibit the classical Turing instability, we have stable kinetics in the absence of diffusion, and thus $\Re(\nu_p) < 0$.  

With the assumption of identical diffusion coefficients, all of the matrices in equation~\eqref{detconfFull} are diagonal and proportional to the identity except for $\bm{M_S}^2$, which can be diagonalised using the eigenvectors of $\bm{J_S}$. Noting that all functions involving $\bm{M_S}$ in \eqref{detconfFull} are both even functions and analytic, we can simultaneously diagonalise the entire condition using these eigenvectors. Doing so, we arrive at the following $n$ scalar conditions from equation~\eqref{detconfFull},
\begin{equation}\label{equalDiffusion}
    c_B m_B\tanh(h m_B)\left(\cosh(\varepsilon m_p)+\frac{c_S}{\eta}m_p\sinh(\varepsilon m_p)\right)+c_S m_p\sinh(\varepsilon m_p)=0,
\end{equation}
where we note that only one of these conditions must be satisfied, and hence they lead to independent roots for $\lambda$. We observe that $\cosh(\varepsilon m_p)=0$ can not occur once $m_p$ does not have zero real part, which is enforced by assumptions. Hence we can divide equation~\eqref{equalDiffusion} by this factor to find,
\begin{equation}\label{equalDiffusion2}
    \frac{c_Sc_Bm_Bm_p}{\eta}\tanh(h m_B)\tanh(\varepsilon m_p)+c_S m_p\tanh(\varepsilon m_p)+c_B m_B\tanh(h m_B)=0.
\end{equation}

We now argue that there are no possible instabilities given  condition \eqref{equalDiffusion2}. We note that for $\Re(\lambda)>0$, both $m_B^2$ and $m_p^2$ must have strictly positive real part. Given $k_q^2\geq0$, $\Re(\lambda)>0$, $\Re(v_p)<0$, we have $\Re(m_p^2)>0$, $\arg(m_p^2)\in (-\pi/2,\pi/2)$ and hence $\arg(m_p)\in(-\pi/4,\pi/4)\cup(-\pi,-3\pi/4)\cup(3\pi/4,\pi)$, with $m_p=0$ excluded (and identical bounds for $m_B$). Thus all roots of $m_p\tanh(\varepsilon m_p)$ and $m_B\tanh(h m_B)$ are excluded and one may divide \eqref{equalDiffusion2} by $c_Sc_Bm_p\tanh(\varepsilon m_p)m_B\tanh(h m_B)$ to obtain,
\begin{equation}\label{equalDiffusion3}
    \frac{1}{\eta}+\frac{1}{c_B m_B\tanh(h m_B)}+\frac{1}{c_S m_p\tanh(\varepsilon m_p)}=0.
\end{equation}

 Letting $z=x+iy \in \mathbb{C}$, $x,y \in \mathbb{R}$, with $\Re(z^2)>0$, we will show that $\Re(z\tanh(z))>0$, and subsequently apply this for $z = m_B$ and $z=m_S$. Letting $Q = (\cosh(x)\cos(y))^2+(\sinh(x)\sin(y))^2 > 0$, we can compute
\begin{equation}\label{ztz}
\Re (z\tanh(z)) = \frac{x\cosh(x)\sinh(x)-y\cos(y)\sin(y)}{Q} = \frac{x\sinh(2x) - y\sin(2y)}{2Q}.
\end{equation}
Additionally, $\Re(z^2)>0$ implies $x^2 > y^2$ which then forces  $\Re(z\tanh(z))>0$ since $x\sinh(2x) - y\sin(2y)>0$ for $x^2>y^2.$
The latter holds given $x>0, ~y\in(-x,x)$  since then
$$ x\sinh(x)>y\sinh(y)\geq y\sin(y).$$
The first inequality holds
as 
the real function  $x\sinh(x)$ is even and, for $x>0$, monotonic increasing, 
while the second inequality follows   using the fact $\sinh(|y|)\geq |y|\geq \sin(|y|)$ for all real $y$ and the odd parity of $y, \sinh(y), \sin(y)$. The case for $x<0$ is then inherited from the case $x>0$ by  parity. 
Hence, the left hand side of equation~\eqref{equalDiffusion3} must have a strictly positive real part, and so this equation can never be satisfied. Therefore, in the case of identical diffusion coefficients within each region, there are no values of $\lambda$ with positive real part. 

\section{Pure diffusion does not induce patterning}
 \label{AppB}
 
We, now, show that   diffusion alone, in the absence of reaction terms,  cannot induce patterning in the modelling framework. In particular in the absence of reaction kinetics, and given our linear interfacial conditions,  each species decouples  and can be considered in isolation. Without loss of generality we consider surface and bulk concentrations of the first species, denoted $c_s$ and $c_B$ below with respective diffusion coefficients $d_B$, $d_S$ in each region. Then, with $\Omega_B$ and $\Omega_S$   denoting the bulk and surface  regions, as in Figure~\ref{Schematic}, and the subscript $t$ denoting a time derivative, we have 
\begin{eqnarray*} \frac 1 2 \frac{\partial}{\partial t}\bigg[\int_{\Omega_B} \mathrm{d}V d_B\nabla c_B\cdot \nabla c_B &+&\int_{\Omega_S} \mathrm{d}V d_S \nabla c_S\cdot \nabla c_S\bigg]
\\ &=&  \int_{\Omega_B} \mathrm{d}V d_B\nabla c_{Bt}\cdot \nabla c_{B} +\int_{\Omega_S} \mathrm{d}V d_S \nabla c_{St}\cdot \nabla c_{S}  \\ &=&
 \int_{\Omega_B} \mathrm{d}V \nabla\cdot( c_{Bt}d_B \nabla c_{B})-c_{Bt}d_B\nabla^2c_B +\int_{\Omega_S} \mathrm{d}V  \nabla \cdot(c_{St} \cdot d_S\nabla c_{S})-c_{St}d_S\nabla^2c_S  \\ &=& \int_{\partial\Omega_B} \mathrm{d}S c_{Bt}d_B \frac{\partial c_{B}}{\partial n} +
  \int_{\partial\Omega_S} \mathrm{d}S c_{St}d_S \frac{\partial c_{S}}{\partial n}
 -\left[\int_{\Omega_B}\mathrm{d}V c_{Bt}^2 + \int_{\Omega_S}\mathrm{d}V c_{St}^2\right]\\
&=&-\left[\int_{\Omega_B}\mathrm{d}V c_{Bt}^2 + \int_{\Omega_S}\mathrm{d}V c_{St}^2\right] \leq0, 
\end{eqnarray*}
where the surface integrals in the fourth line vanish, courtesy of the zero flux boundary conditions (\ref{BC1}) and (\ref{BC2}), and the  interfacial conditions (\ref{coupling}), on noting relations  (\ref{geqn}) and (\ref{sceqn}). This can also be recognised as a free energy inequality, or equivalently an entropy inequality, corresponding to the second law of thermodynamics given a Fickian diffusive flux as a constitutive relation \cite{gurtin}.  Hence a standard measure of heterogeneity   cannot increase and thus   initial conditions that are close to homogeneous (in the sense of a suitable Sobolov norm) cannot induce patterns.

\bibliographystyle{abbrv}
\bibliography{refs}

\end{document}